\documentclass{aa}  

\usepackage{graphicx}
\usepackage{txfonts}
\usepackage{lipsum}
\usepackage{subcaption}         % necessary for continued figures, example in section 3
\usepackage{lscape}             % to rotate a single page table, example in appendix.
\usepackage{placeins}           % useful with \FloatBarrier, to keep 
\makeatletter
\@fleqnfalse
\makeatother   

\usepackage{booktabs}
\usepackage{makecell}
\usepackage{nicematrix}
\usepackage{pifont}
\usepackage{changepage}
\usepackage{pdflscape}
\usepackage{multirow}
\usepackage{multicol}
\usepackage{CJKutf8}

\usepackage{hyperref}
\begin{document}

   \title{Drift-dominated dust evolution}
   \subtitle{A multiwavelength analysis of dust rings in CY Tau and DoAr25}

%%%%%%%%%%%%%%%%%%%%%%%%%%%%%%%%%%%%%%%%
% Please do not include ORCIDs next to author names.
% Only ORCIDs authenticated by individual authors in EDP Sciences editorial system will be taken into account.
% ORCIDs included here will be removed.
%%%%%%%%%%%%%%%%%%%%%%%%%%%%%%%%%%%%%%%%

   \author{Anastasia Topalidou\inst{\ref{inst1}\fnmsep\ref{inst2}}\thanks{Corresponding author: A.Topalidou.pgr@leeds.ac.uk}
   \and Nienke van der Marel\inst{\ref{inst1}}
   \and Paola Pinilla\inst{\ref{inst3}}
   \and Haochang Jiang (\begin{CJK*}{UTF8}{gbsn}蒋昊昌\end{CJK*})\inst{\ref{inst4}} 
   \and Osmar M. Guerra-Alvarado\inst{\ref{inst1}}
   \and Brodie J. Norfolk\inst{\ref{inst5}}}

   \institute{Leiden Observatory, Leiden University, Leiden, The Netherlands \label{inst1}
   \and School of Physics and Astronomy, University of Leeds, Leeds, LS29JT, UK \label{inst2}
   \and Mullard Space Science Laboratory, University College London, Dorking, UK \label{inst3}
   \and Max-Planck-Institut für Astronomie, Königstuhl 17, 69117 Heidelberg, Germany\label{inst4}
   \and Centre for Astrophysics and Supercomputing (CAS), Swinburne University of Technology, Hawthorn, Victoria 3122, Australia \label{inst5}}

   \date{Received 4 June 2026 / Accepted 6 August 2026}

% \abstract{}{}{}{}{}
% 5 {} token are mandatory
 
  \abstract
   {Protoplanetary disks are the birthplaces of planets, and understanding the distribution and evolution of dust within them is essential for tracing the early stages of planet formation.}
   {Following the DSHARP survey, we aim to constrain the radial distribution of the dust surface density and the maximum emitting grain size in the disks of CY Tau and DoAr 25 and assess the relative importance of dust trapping and radial drift in shaping the dust population in disks.} 
   {We present multiwavelength continuum observations of CY Tau and DoAr~25 from ALMA at 2.4 and 1.25 mm and archival VLA data at 8.8 mm. Combining the observations with dust continuum forward modeling, we sought to infer a radial profile of the two systems of the dust surface density and maximum emitting grain size.}
   {For DoAr~25, the results reveal a prominent dust ring at 111 AU and evidence for large (centimeter-sized) grains extending to the outer disk. In contrast, CY Tau exhibits a more compact and smoother structure, with both the dust surface density and grain size decreasing monotonically with radius. A tentative shoulder feature in the radial profiles, located between 20--40 AU, may point to a marginal substructure, possibly a weak dust trap.}
   {DoAr 25 shows evidence for a dust trap capable of halting inward drift, while CY Tau exhibits behavior consistent with evolution dominated by radial drift. Together, these results highlight how dust evolution processes shape the disk morphology and grain-size distributions, even in the low-mass stellar regime.}


   \maketitle
    \nolinenumbers

%%%%%%%%%%%%%%%%%%%%%%%%%%%%%%%%%%%%%%%%%%%%%%%%%%%%%%%%%%%%%%
\section{Introduction}
Until a decade ago, protoplanetary disks  were thought to be smooth, axisymmetric structures whose evolution was governed by radial drift \citep{whipple1973radial,weidenschilling1977aerodynamics,gorti2009time,macias2018multiple} and fragmentation of dust grains \citep{blum2008growth}. Due to the advent of the Atacama Large Millimeter/submillimeter Array (ALMA), our resolving power has grown to 0.02-0.03" in angular resolution, which corresponds to a few AU in spatial resolution in linear scale for nearby disks. This led to the discovery of fine structures (sets of concentric bright rings and dark gaps) in the protoplanetary disk surrounding the young star HL Tau which was imaged using the 2014 ALMA Long Baseline Campaign \citep{brogan20152014} and numerous other disks since then \citep[e.g.,][]{andrews2020observations}.

The grain distribution in disks was originally believed to follow a radial size, sorting in the absence of substructures: their distribution would be based on their size as larger grains experience greater aerodynamic drag and radially migrate faster \citep{Tanaka2005,Laibe2008,Birnstiel2012}. Early multiwavelength millimeter observations indeed showed disk images traced at longer wavelength observations (probing larger grains) were more compact, for example DoAr 25 and CYTau \citep{perez2015grain} as well as CQ Tau \citep{trotta2013constraints}. However, newer high spatial resolution comparisons have revealed that for several other disks, dust grains of varying sizes exhibit extended emission that is located at similar radial locations in the disk (see \citet{macias2021characterizing, tazzari2021multiwavelength}).
High resolution surveys have revealed that substructures are ubiquitous at least in bright disks \citep{pinilla2018homogeneous, Long2019,van2021stellar,bosschaart2026gaps}, and they are generally less common in smaller disks \citep{Guerra2025}, although this result could be biased due to the available resolution.

The key to understanding the origin and role of the substructures in disks is their dust content \citep{2021A&A...648A..33M}. Notably, if dust grains accumulate at gas pressure bumps (local pressure maxima), these regions can act as dust traps. Various physical processes can result in the presence of traps in disks, such as condensation fronts \citep{Zhang2015}, self-induced dust traps \citep{Gonzalez2015,Gonzalez2017}, or nonideal magnetohydrodynamic (MHD) effects \citep{Ueda2019}, but usually planet-disk interactions are regarded as the main driving force \citep{Dong2015, Lodato2019}. Gas pressure bumps can halt the radial drift of large particles, allowing them to grow up to planetesimal sizes, therefore enabling the planet formation process \citep{macias2019characterization}. 

This raises a fundamental question in planet formation theory, namely whether dust traps are necessary precursors to planet formation or are instead created as a consequence of embedded planets interacting with the disk. This presents a classic chicken-and-egg problem: pressure bumps can halt radial drift and allow solids to grow into planetesimals, yet the very presence of planets can give rise to these pressure maxima. Disentangling the origin of disk substructures is essential to understanding whether they are signs of active planet formation or its enabling conditions. By probing the emission at different wavelengths, we can trace dust grains of various sizes, since the dominant emitting grain size scales as $\sim \lambda/2\pi$ \citep{draine2006submillimeter}. 

Substructures such as rings, gaps, and cavities can strongly affect the interpretation of millimeter- and centimeter-wavelength observations, as optically thick regions or unresolved substructures may bias estimates of the dust surface density and the apparent grain growth. Multiwavelength analysis of substructures in disks \citep[e.g.,][]{macias2021characterizing} have primarily focused on disks around solar-type stars. However, dust evolution models predict distinct differences in the efficiency of drift and grain growth in pressure bumps in disks surrounding different types of stars \citep{2020A&A...635A.105P}. In this context, disks such as CY Tau and DoAr 25 surrounding lower-mass stars  (0.3 and 0.65 $M_\odot$ respectively) provide ideal laboratories to study how substructures manifest across wavelengths and to assess their impact on the measured dust properties, offering insight into both the efficiency of grain growth and the potential role of dust traps in low-mass disks.

To build on the sample of disks with multiwavelength observations, we present new multiwavelength ALMA observations of DoAr~25 and CY Tau, combined with a reanalysis of previously published Very Large Array (VLA) data. These new observations offer an opportunity to probe more optically thin emission and test the hypothesis that dust substructures are present and significant in these disks. The protoplanetary disks surrounding the young stars CY Tau and DoAr 25 are located, respectively, in the Taurus star-forming region at a distance of 125.3 pc \citep{2020yCat.1350....0G} and in the L1688 dark cloud at a distance of 138.2 pc \citep{2020yCat.1350....0G}. CY Tau and DoAr 25 are pre-main-sequence stars of spectral types M2V/M1.5 \citep{antonellini2016mid} and K5 \citep{andrews2009protoplanetary}, respectively. Both are relatively young: CY Tau has an estimated age of 2--3\,Myr \citep{bertout2007evolution, guilloteau2014masses}, while DoAr 25 is approximately 4\,Myr old \citep{andrews2009protoplanetary}. Both stars exhibit significant emission excess above the stellar photosphere from near-infrared to millimeter wavelengths \citep{olofsson2009c2d, mcclure2010evolutionary}. DoAr 25 has been imaged at submillimeter wavelengths with the Submillimeter Array (SMA) \citep{andrews2008structure, andrews2009protoplanetary, perez2015grain}, at 2.8 mm using The Combined Array for Research in Millimeter-wave Astronomy (CARMA) and at 8.0, 9.8, and 50 mm using the VLA. It has also been observed in scattered light at 1.6 $\mu$m with the Spectro-Polarimetric High-contrast Exoplanet REsearch instrument (SPHERE) on ESO's Very Large Telescope (VLT) showing complex vertical settling \citep{antilen2026diverse}. CY Tau was observed at 1.3 and 2.8 mm with the Plateau de Bure Interferometer \citep{guilloteau2011dual} and CARMA, and at 7.1 and 50 mm using the VLA \citep{perez2015grain}.

For this study, we analyzed dust continuum emission at 1.25 and 2.24 mm of CY Tau and DoAr 25. First, we assessed whether the observed substructures arise primarily from grain growth or from planet-disk interactions. We also determined the maximum emitting grain size and dust surface density. Current estimates of protoplanetary disk dust masses based on the millimeter flux appear inadequate to account for the solid mass budget required for the observed exoplanet population \citep{greaves2010have, manara2018protoplanetary,mulders2021}, suggesting that by the Class II stage, much of the solid material may have already been incorporated into larger bodies or planets. This discrepancy could be explained if a significant portion of the mass is already in large bodies that millimeter observations do not fully capture, as suggested by \citet{najita2014mass}. In addition, optical depth effects can further bias these measurements, as regions that are optically thick may hide a substantial fraction of the solid material from the observed flux \citep{2025A&A...694A.290G}. Consequently, measuring the spatially resolved surface density of solids at multiple wavelengths is crucial, as it allows us to disentangle optical depth effects and obtain a more accurate estimate of the dust mass available for the formation of rocky planets.

The observational methods and results are presented in Section \ref{sec:observations}. The visibility fitting, multiwavelength analysis, and dust continuum modeling are presented in Section \ref{sec:analysis}. The discussion of our results, including a comparison with a broader sample of disks is presented in Section \ref{sec:discussion}, followed by our conclusions in Section \ref{sec:cnclusions}.

\section{Observations}\label{sec:observations}

This work makes use of new continuum observations obtained through ALMA program 2022.1.01284.S (PI: B. Norfolk), observed between May 2023 and September 2023. The data set includes Band 4 data for DoAr 25 and CY Tau, and additional Band 6 observations at high angular resolution for CY Tau. In addition, low-resolution data was used from project 2022.1.01302.S (PI: G. Mulders), observed from January 2023 to June 2023 also presented in \citet{bosschaart2026gaps}. The continuum was constructed by averaging line-free channels from four spectral windows spanning 126.2--141.7\,GHz (with bandwidths 0.94 and 1.87 GHz), resulting in an effective central frequency of 134\,GHz. The Band~6 continuum was similarly obtained from four spectral windows covering 214.9--233.9\,GHz (with a bandwidth of 2 GHz), resulting in an effective central frequency of 231.5\,GHz. The baseline range, calibrators, integration time, resulting angular resolution and observation date of each execution block are listed in Table~\ref{tab:data}. The data were reduced using the provided pipeline script with CASA version 6.4.1.12 (Common Astronomy Software Applications; \cite{bean2022casa}). The calibrated visibility data were first reduced by channel averaging and time-binning over 30-second intervals to produce a more manageable dataset.

Imaging was performed using the \texttt{tclean} task in CASA version 6.6.6, using a Briggs weighting scheme with a robust parameter of 0.5. The image cell size and image dimensions were selected to ensure proper sampling of the synthesized beam (3-5 pixels across the beam), and to encompass the full spatial extent of the disk.

\begin{table*}[htbp!]
    \centering
    \caption{Observational details of the new ALMA observations used in this work for CY Tau and DoAr 25.}
    \begin{tabular}{ccccccc}
    \hline \hline
    Target & Band & Ang. Res. ["] & Baselines [m] & Calibrators & Int. Time [m] & Obs. Date \\
    \hline
    CY Tau & Band 4 & 0.289 & 92-888 &  J0403+2600, J0359+2758 & 34.3 & 03-05-2023\\
          &        & 0.043 & 719-5944 & J0429+2724, J0433+2905 & 70.8 & 12-08-2023\\
          & Band 6 & 0.124 & 27-3638 & J0438+3004, J0440+2728 & 14.5 & 15-09-2023\\
          &        & 0.048 & 279-3284 & J0433+2905, J0429+2724 & 23.3 & 31-05-2023\\
    \hline
    DoAr 25 & Band 4 & 0.178 & 151-1490 & J1633-2557, J1617-2537 & 9.1 & 30-05-2023\\
           &        & 0.033 & 988-8010 & J1617-2537 & 41.1 & 06-08-2023\\

     \hline
    \end{tabular}
    \tablefoot{For the angular resolution Briggs weighting was used with robust=0.5.}
    \label{tab:data}
\end{table*}

We then proceeded to self-calibrate the datasets. The self-calibration was primarily focused on the Band 6 data for CY Tau, as higher-frequency observations are more sensitive to atmospheric phase fluctuations. However, we also applied self-calibration to the Band 4 data to maximize the signal-to-noise ratio (S/N) in each image. The phase calibration process began by setting \texttt{savemodel=True} in the \texttt{tclean} command and performing a shallow clean (10 iterations) to save the model. After measuring the properties of the initial image, we identified the antenna closest to the array center using the \texttt{plotants} command. We then ran \texttt{gaincal}, first with a solution interval of \texttt{inf} to generate the initial calibration table, followed by \texttt{applycal} to apply the corrections. Then we split the calibrated data from the \texttt{corrected} column. This newly calibrated dataset was cleaned again using a $3\sigma$ threshold, where $\sigma$ is the RMS of the first image. These steps were repeated iteratively, reducing the solution interval from \texttt{inf} to 60s and then 18s, while monitoring changes in peak intensity, RMS noise, and beam size. The process was terminated when the peak intensity and RMS plateaued, typically after 4-6 rounds (e.g., for CY Tau in Band 6 we can see that the S/N during the first round is 35 which increases to 150 by the end of the self-calibration process). Amplitude calibration was tested on all datasets but was ultimately adopted for only one, as it degraded the beam quality in the others. Notably, in early attempts with self-calibration, the beam size increased during calibration, resulting in a loss of resolution. To counter this, we used the \texttt{applymode='calonly'} option to prevent this degradation.

After the self-calibration was completed we combined the short and long baseline datasets, using the CASA command \texttt{concat} so we can acquire an even higher S/N provided from the short baselines all while keeping the details provided by the long baselines. For the combination, we used a frequency tolerance of 10MHz and a direction tolerance of 0.1 arcsec to ensure that in the final measurement set all the data remained in the same field. The properties before and after the combination are listed in Table~\ref{tab:properties_comb}

\begin{table*}[htbp!]

    \centering
    \caption{Calculated image properties of both disks.}
    \renewcommand{\arraystretch}{1.2}
    \begin{tabular}{lccccccc}
    \hline\hline
\multirow{2}{*}{Property} & \multicolumn{3}{c}{DoAr 25} & \multicolumn{3}{c}{CY Tau} \\
     & B4 & B6 & Ka  & B4 & B6 & Ka  \\
    
\midrule
Flux density [$\times 10^{-2}$ Jy] & 6.99 & 23.9 & 10.2 & 4.03 & 11.27 & 0.17\\
Peak [$\times 10^{-4}$ Jy] & 6.55 & 32.0 & 0.14 & 13.5 & 30.58 & 5.0\\
RMS [$\times 10^{-6}$ Jy]  & 8.49 & 72.6 & 0.12 & 6.65 & 18.74 & 0.34\\
% SNR &  77.1 & 44 & 117 & 203.0 & 147.8 & 1407.8 \\
\multirow[t]{2}{*}{Beam} & 0.038 & 0.041 & 0.099 & 0.06 &0.079 & 0.175 \\
{["]}& $\times$ 0.033& 0.022 & 0.047 & $\times$0.041 & $\times$0.041& 0.122\\
    \hline
    \end{tabular}
    \label{tab:properties_comb}
\\ \textbf{Notes.} A Briggs weighting scheme with a robust parameter of 0.5 was used to create the images.
\end{table*}

\begin{figure*}[!htbp]
    \centering
    \includegraphics[width=\textwidth]{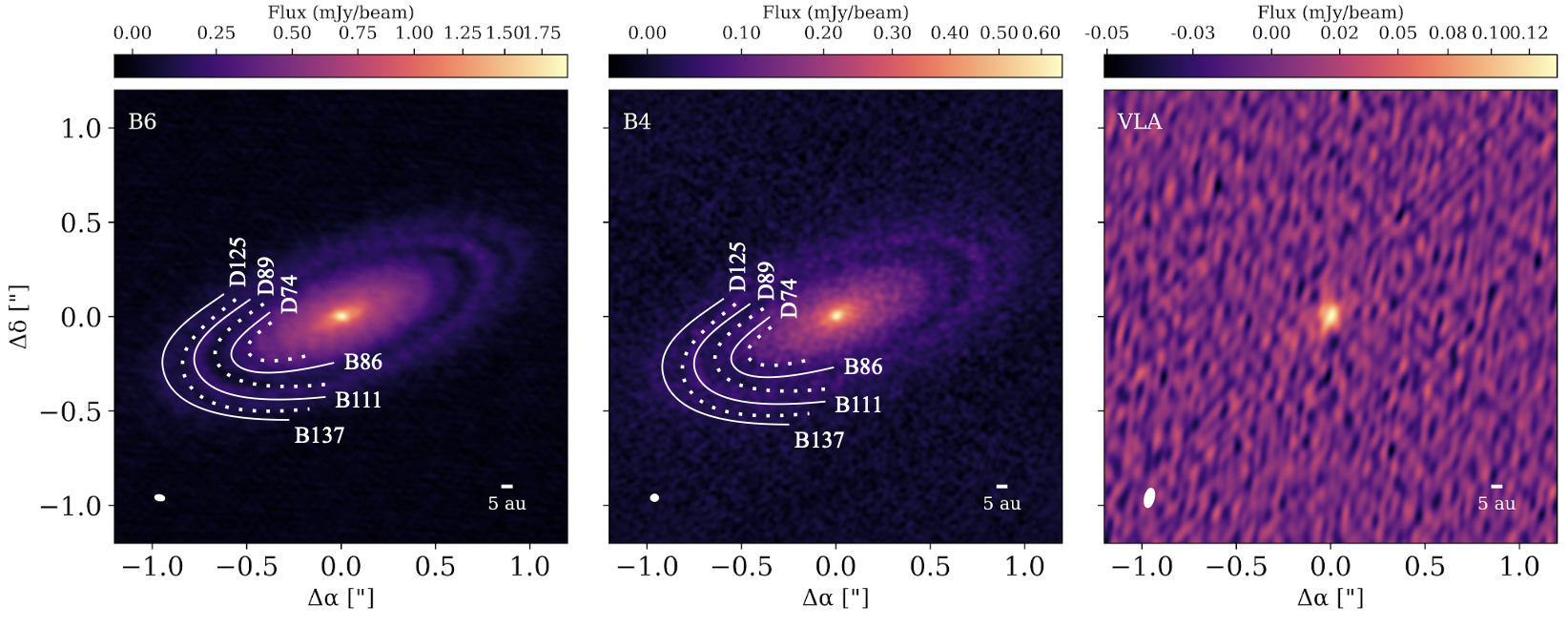}
    \caption{Continuum images of DoAr 25 at multiple wavelengths, from left to right: Band 6 (1.25 mm), Band 4 (2.24 mm), and VLA (8.8 mm). The new Band 4 image shows significantly more extended emission, similar to the Band 6 data, compared to the VLA image. Bright rings are labeled "B" and dark gaps are labeled "D" following the convention of \cite{huang2018disk}. The beam size is shown with the ellipse in the bottom-left corner, and a 5 AU scale bar is included to indicate the physical scale. An arcsinh stretch is applied to the color scale of each panel to make substructures in the outer regions more visible.}
    \label{fig:images_doar}
\end{figure*}

\begin{figure*}[!htbp]
    \centering
    \includegraphics[width=\textwidth]{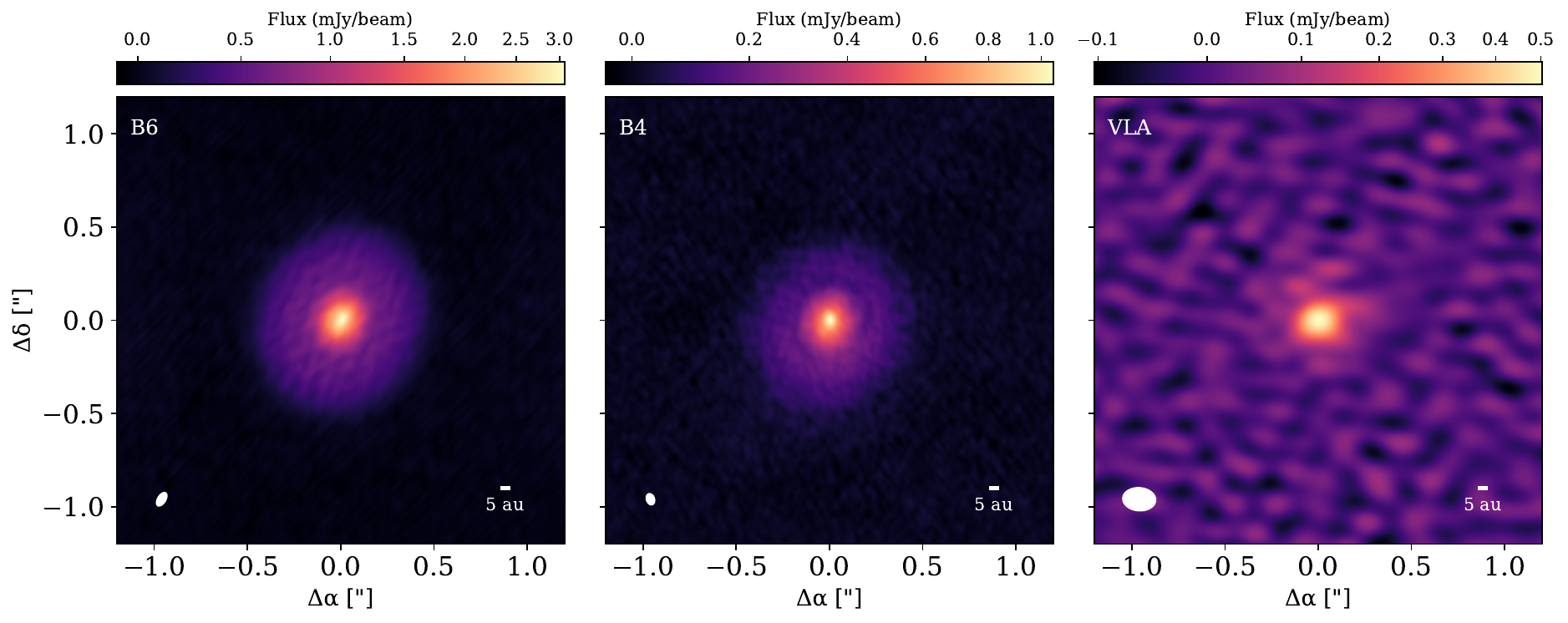}
    \caption{Continuum images of CY Tau at multiple wavelengths, from left to right: Band 6 (1.25 mm), Band 4 (2.24 mm), and VLA (8.8 mm). The new Band 4 and Band 6 data show significantly more extended emission compared to the VLA data. The beam size is shown with the ellipse in the bottom-left corner, and a 5 AU scale bar is included to indicate the physical scale. An arcsinh stretch is applied to the color scale of each panel to make substructures in the outer regions more visible.}
    \label{fig:images_cytau}
\end{figure*}

We present the results of the self-calibration and concatenation procedures performed on the new observational data, in Figs.~\ref{fig:images_doar}--\ref{fig:images_cytau}. The newly produced ALMA images at 1.3 and 2.2 mm are shown alongside archival VLA data at 8.8 mm \citep{perez2015grain} and, in the case of DoAr 25, ALMA Band 6 (1.3 mm) data from the DSHARP survey \citep{andrews2018disk}. 

Fig.~\ref{fig:images_doar} shows the continuum images of DoAr~25. The new Band 4 data show similar disk structure and morphology as observed in the existing Band 6 images. Both Band 4 and 6 continuum images show significantly more extended emission than in the VLA continuum. A prominent ring at 111 AU is visible.

CY Tau (Fig.~\ref{fig:images_cytau}) shows a similar pattern as seen in the continuum images of DoAr 25. The new Band 4 and Band 6 images reveal significantly more extended emission compared to what is observed in the VLA data. The Band 6 data show primarily a smooth disk structure. In the Band 4 data, a faint gap is detected at approximately 25 AU radius.

\section{Analysis}\label{sec:analysis}

\subsection{Radial brightness profiles}

\begin{figure}[htbp!]
    \centering

    \begin{subfigure}{\linewidth}
        \centering
        \includegraphics[width=\linewidth]{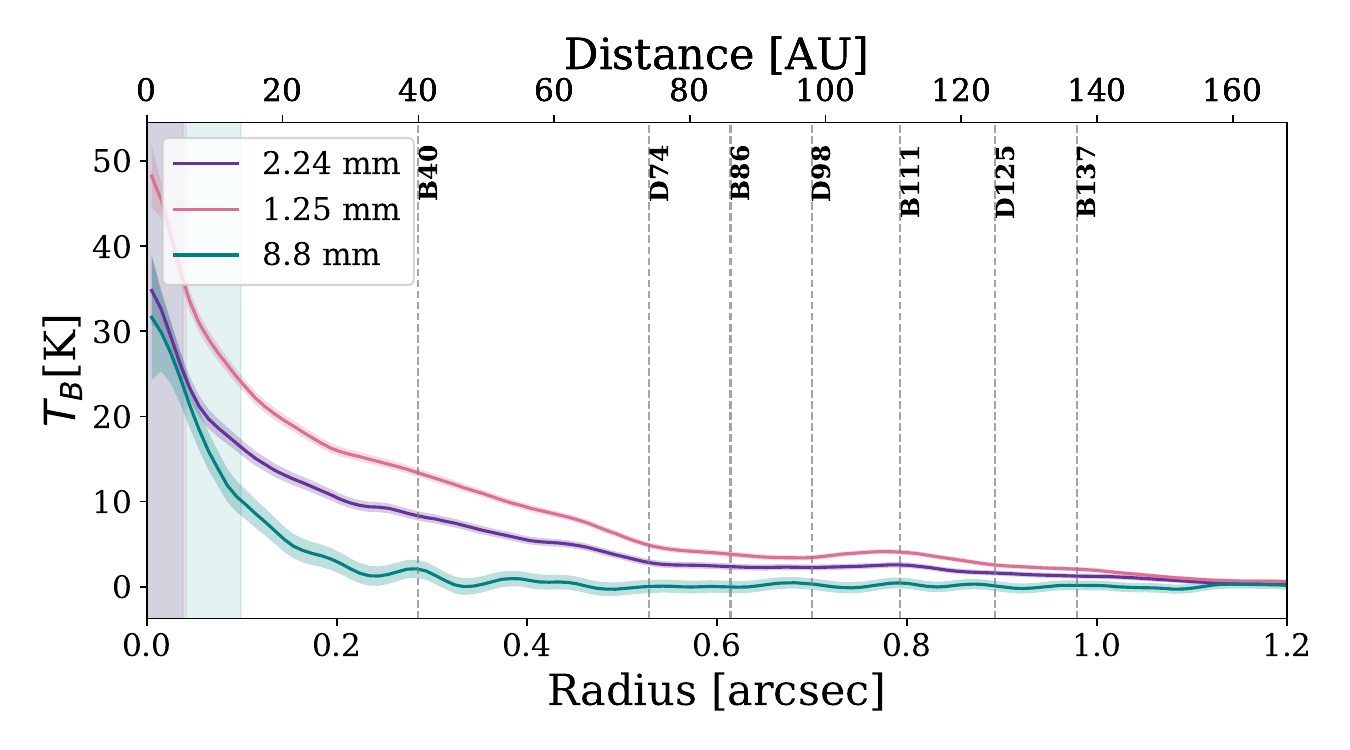}
        \caption{DoAr 25.}
        \label{fig:radial_prof_doar25}
    \end{subfigure}

    \begin{subfigure}{\linewidth}
        \centering
        \includegraphics[width=\linewidth]{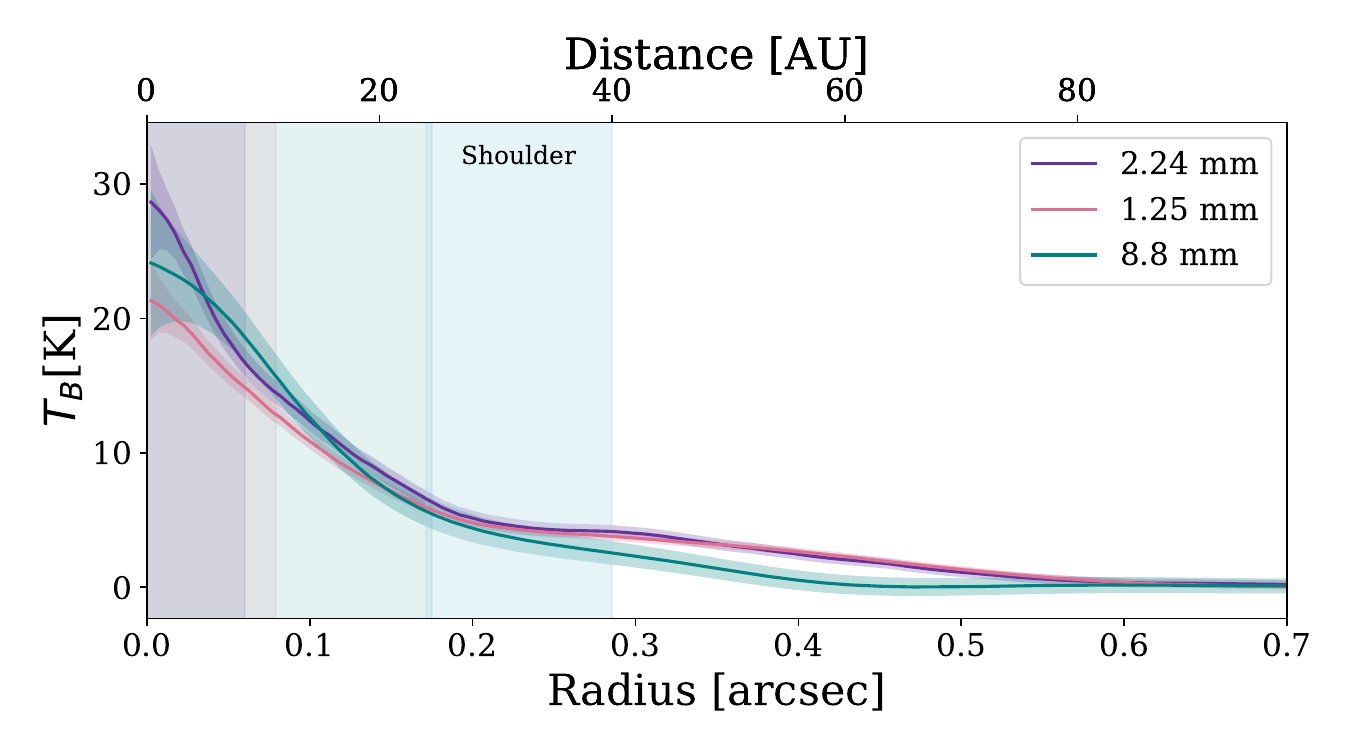}
        \caption{CY Tau.}
        \label{fig:radial_prof_cytau}
    \end{subfigure}

    \caption{Radial brightness temperature profiles of DoAr 25 (top) and CY Tau (bottom) at multiple wavelengths: Band 6 (purple), Band 4 (pink), and VLA (teal). The prominent ring at 111 AU in DoAr 25 is clearly visible in all datasets, while a tentative ring at 40 AU is visible in the VLA dataset. CY Tau exhibits a predominantly smooth disk. A shoulder feature is visible between 24-40 AU, noted with light blue, appearing as a brightness enhancement in the profile. This feature corresponds to the subtle substructure observed in the continuum images. Shaded regions indicate the error of the mean at each radius. The shaded regions in the beginning of the plots correspond to the resolution of each dataset.}
    \label{fig:radial_profiles_combined}
\end{figure}

To carry out the multiwavelength analysis of our data, it is essential to obtain a well-defined representation of the radial brightness profiles. These profiles were derived by azimuthally averaging the continuum images, that accounts for both the disk inclination and position angle, $i=62^\circ \ \text{and} \ PA=109^\circ$ for DoAr~25 and $i=28^\circ \ \text{and} \ PA=63^\circ$ for CY Tau \citep{perez2015grain}. This method operates on the two-dimensional image plane, where a series of concentric annular rings centered on the source are constructed. The mean intensity within each ring is then computed to produce a deprojected brightness profile as a function of radius. The width of each ring is set to one-third of the synthesized beam size, ensuring that the resulting radial bins are statistically independent\footnote{Adapted from the \texttt{Radial-Profile} python code by \href{https://github.com/emaciasq}{Enrique Macías}}. The uncertainty of the radial profiles is computed as the error of the mean at each radius, considering the beam size as the smallest independent unit of area

\begin{equation*} 
\sigma_{\bar{I}}\,{=}\,\frac{\sigma_i}{\sqrt{N_{\textrm{B}}}}\,{=}\,\frac{\sigma_i}{\sqrt{A_i/A_{\textrm{beam}}}}, 
\end{equation*}

where $\sigma_i$ is the standard deviation within the concentric ellipse, $N_B$ is the number of beams within the ellipse, $A_i$ is the area of the ellipse, and $A_{beam}$ is the area of the beam. In Fig.~\ref{fig:radial_profiles_combined} we present the brightness temperature profiles derived from the aforementioned radial profiles with circular beams corresponding to the major axis of each dataset. In the profile of DoAr 25 in Fig.~\ref{fig:radial_prof_doar25}, we indicate bright rings with "B" and dark gaps with "D", following the convention established in \cite{huang2018disk}. The most significant ring remains located at 111 AU, but there are inflections points in the VLA profile within 60 AU. The most prominent one is at 40 AU, tentatively given the lower S/N of the VLA observations. In the profile of CY Tau in  Fig.~\ref{fig:radial_prof_cytau} we observe a brightness profile bump that can be characterized as a shoulder feature spanning 24-40 AU. The shaded areas at the beginning of the plots indicate the resolution of each dataset.

\subsection{Multiwavelength dust continuum forward modeling}\label{sec:rad_transfer}
We aim to compute the key physical dust properties as function of radius: the dust temperature $T_d(r)$, surface density $\Sigma(r)$ and maximum grain size $a_{\text{max}}(r)$. We compute these radial profiles following the methodology presented by \citet{carrasco2019radial} and \citet{macias2021characterizing}. The input consists of de-projected radial intensity profiles extracted at three different wavelengths: 1.25~mm, 2.4~mm, and 8.8~mm, using the azimuthal averaging technique described previously. These are modeled using a multiwavelength dust continuum forward model that assumes a slab geometry, i.e., the disk is treated locally as a plane-parallel, isothermal slab of dust that radiates thermally and scatters radiation \citep{sierra2019analytical}. One caveat of this model is that the vertical emitting height is assumed to be the same for all wavelengths.

The formalism is based on
\begin{equation}
    I_{\nu} = S_{\nu} \left(1 - e^{-\tau_{\nu}}\right),
\label{eq:rad_transfer}
\end{equation}
where $I_{\nu}$ is the emergent specific intensity, $S_{\nu}$ is the source function incorporating scattering, and $\tau_{\nu}$ is the total optical depth
\begin{equation}
    \tau_{\nu} = \Sigma \left( \kappa_{\nu} + \sigma_{\nu} \right),
\end{equation}
with $\kappa_{\nu}$ and $\sigma_{\nu}$ denoting the absorption and scattering opacities, respectively.

To model the contribution of scattering to the emergent intensity, we adopted the two-stream approximation from \citet{carrasco2019radial}, yielding
\begin{equation}
    I_{\nu} = B_{\nu}(T_d) \left[1 - e^{-\tau_{\nu}/\mu} + \omega_{\nu} F(\tau_\nu, \omega_\nu) \right],
\end{equation}
where $B_{\nu}(T_d)$ is the Planck function, $\mu = \cos(i)$ accounts for the disk inclination $i$ (a face-on disk corresponds to i=0$^\circ$), $\omega_\nu = \kappa^{\text{sca}}_\nu / (\kappa^{\text{abs}}_\nu + \kappa^{\text{sca}}_\nu)$ is the albedo, and $F(\tau_\nu, \omega_\nu)$ is a correction factor for isotropic scattering, which is particularly relevant in moderately optically thick regimes. It is defined as 

\begin{align}
F(\tau_{\nu},\omega_{\nu})=&\frac{1}{\textrm{exp}(-\sqrt{3}\epsilon_{\nu}\tau_{\nu})(\epsilon_{\nu}-1)-(\epsilon_{\nu}+1)}\nonumber\\ &\times\,\left[\frac{1-\textrm{exp}(-(\sqrt{3}\epsilon_{\nu}+1/\mu)\tau_{\nu})}{\sqrt{3}\epsilon_{\nu}\mu+1}\right.\\ &\left.+\,\frac{\textrm{exp}(-\tau_{\nu}/\mu)-\textrm{exp}(-\sqrt{3}\epsilon_{\nu}\tau_{\nu})}{\sqrt{3}\epsilon_{\nu}\mu-1}\right],\nonumber 
\end{align}
where $\epsilon_\nu = \sqrt{1-\omega_\nu}$.

The opacities we adopt in this work are the DSHARP opacities by \citet{birnstiel2018disk}\footnote{The python package  \href{https://github.com/birnstiel/dsharp_opac/tree/master}{\texttt{dsharp\_opac}}}. They assume particles without porosity and a composition of 60\% silicates, 15\% troilite and 25\% refractory organics by mass, with optical constants compiled from laboratory experiments, with a minimum grain size of $10^{-5}$ cm. The model computes size-averaged absorption and scattering opacities for compact, spherical grains following a truncated power-law size distribution, and accounts for the full Mie theory including scattering effects. For our code we interpolated the opacities from the precomputed tables using \texttt{RegularGridInterpolator}, as functions of both wavelength $\lambda$ and grain size $a_{\text{max}}$, to construct $\kappa_{\nu}(a_{\text{max}}, \lambda)$.

We apply a Bayesian approach to infer the disk parameters using Markov Chain Monte Carlo (MCMC), with the \texttt{emcee} package \citep{2013PASP..125..306F}. For each radial bin, three parameters are fitted simultaneously: the midplane temperature $T_d$, the surface density $\Sigma$ (in log scale), and the maximum grain size $a_{\text{max}}$ (also in log scale). The prior on $T_d$ is derived from the expected profile for a passively irradiated, flared disk in radiative equilibrium \citep{chiang1997spectral}:
\begin{equation}
    T_{\text{mid}}(r) = \left( \frac{\phi L_*}{8 \pi \sigma_{\text{SB}} r^2} \right)^{1/4},
\end{equation}
where $\phi$ is the disk flaring angle and $L_*$ is the stellar luminosity. We assume a uniform distribution for $\phi$ between 0.01 and 0.06 \citep{huang2018disk}, and a normal distribution for $L_*$ centered at $0.95\,L_{\odot}$ for DoAr~25 \citep{villenave2025turbulence} and $0.4\,L_{\odot}$ for CY Tau \citep{antonellini2016mid} with a 15\% standard deviation. The surface density $\Sigma(r)$ is assigned a flat prior in logarithmic space with physical bounds, and the prior for $a_{\text{max}}$ is uniform in log scale. Additionally, we assume that the grains follow a power-law distribution in the disk

\begin{equation}
    n(a) \propto a^p ; a_{min}< a < a_{max},
\end{equation}
where we set p=-3.5, the typical ISM value \citep{mathis1977size} and $a_{min}\ \text{and}\ a_{max}$ are minimum and maximum emitting grain sizes in the disk.

However, due to the limited wavelength coverage of our data, a degeneracy arises between temperature and maximum grain size. To help mitigate this, we split the modeling into two separate regimes: one with a grain size prior spanning $1\,\mu\text{m} \leq a_{\text{max}} \leq 1\text{mm}$, and another covering $1\text{mm} \leq a_{\text{max}} \leq 10\,\text{cm}$. This approach helps distinguish between models dominated by small versus large grains.

Even with this separation, degeneracies persisted-particularly between temperature and grain size-so we adopted a second fitting strategy in which the temperature profile was fixed to a power-law form, which closely follows what the first modeling approach showed:
\begin{equation}\label{eq:power_law}
    T(r) = T_0 \left( \frac{R}{R_0} \right)^{-q},
\end{equation}
with $q = 0.5$, $R_0 = 10\,\text{AU}$, and $T_0 = 100\,\text{K}$, consistent with theoretical expectations for passive disks \citep{chiang1997spectral, macias2019characterization}. This allowed us to isolate the effects of grain growth and surface density more robustly.

The likelihood is defined using a Gaussian log-likelihood function:
\begin{equation}
    \ln \mathcal{L} = -0.5 \sum_i \left( \frac{I_{\nu, \text{obs}}(r_i) - I_{\nu, \text{mod}}(r_i)}{\sigma_{\nu}(r_i)} \right)^2,
\end{equation}
where $I_{\nu, \text{obs}}$ and $I_{\nu, \text{mod}}$ represent the observed and modeled intensities, and $\sigma_{\nu}$ is the uncertainty at each radial bin.

We use the \texttt{emcee} package-an affine-invariant ensemble sampler developed by \citet{2013PASP..125..306F}-with 100 walkers and 1000 burn-in steps, followed by 1000 steps for posterior sampling. Parameter initialization is drawn from the prior distributions, and autocorrelation times are tracked to ensure convergence.

We present the results from the two modeling approaches: %outlined in Section \ref{sec:rad_transfer}:
\begin{enumerate}
    \item  Fitting all three properties (temperature, dust surface density, and maximum emitting grain size) simultaneously,
    \item Applying a fixed power law for the temperature prior and fitting only the dust surface density and maximum emitting grain size.
\end{enumerate}

In Figs.~\ref{fig:doar_large_prior} and \ref{fig:doar_small_prior} we can see how, when the temperature is a free parameter, the sampler struggles to efficiently fit the dust surface density and maximum emitting grain size, clearly indicating the necessity of implementing a fixed power law distribution for the temperature prior.

\begin{figure}[htbp!]
    \centering
        \begin{subfigure}{\linewidth}
        \centering
    \includegraphics[width=\linewidth]{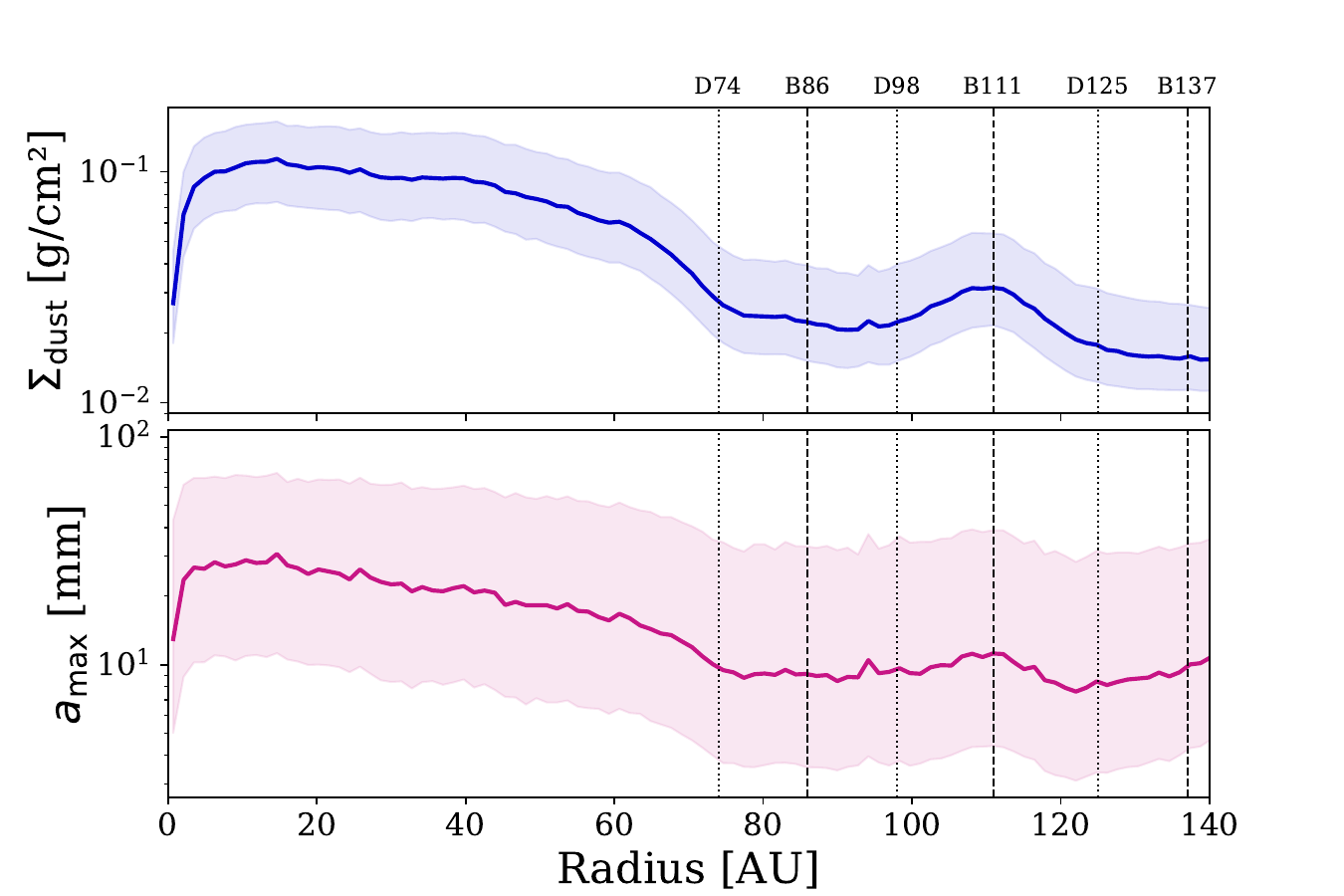}
    \caption{DoAr 25}
    \label{fig:properties_doar25}
  \end{subfigure}

    \begin{subfigure}{\linewidth}
    \centering
    \includegraphics[width=\linewidth]{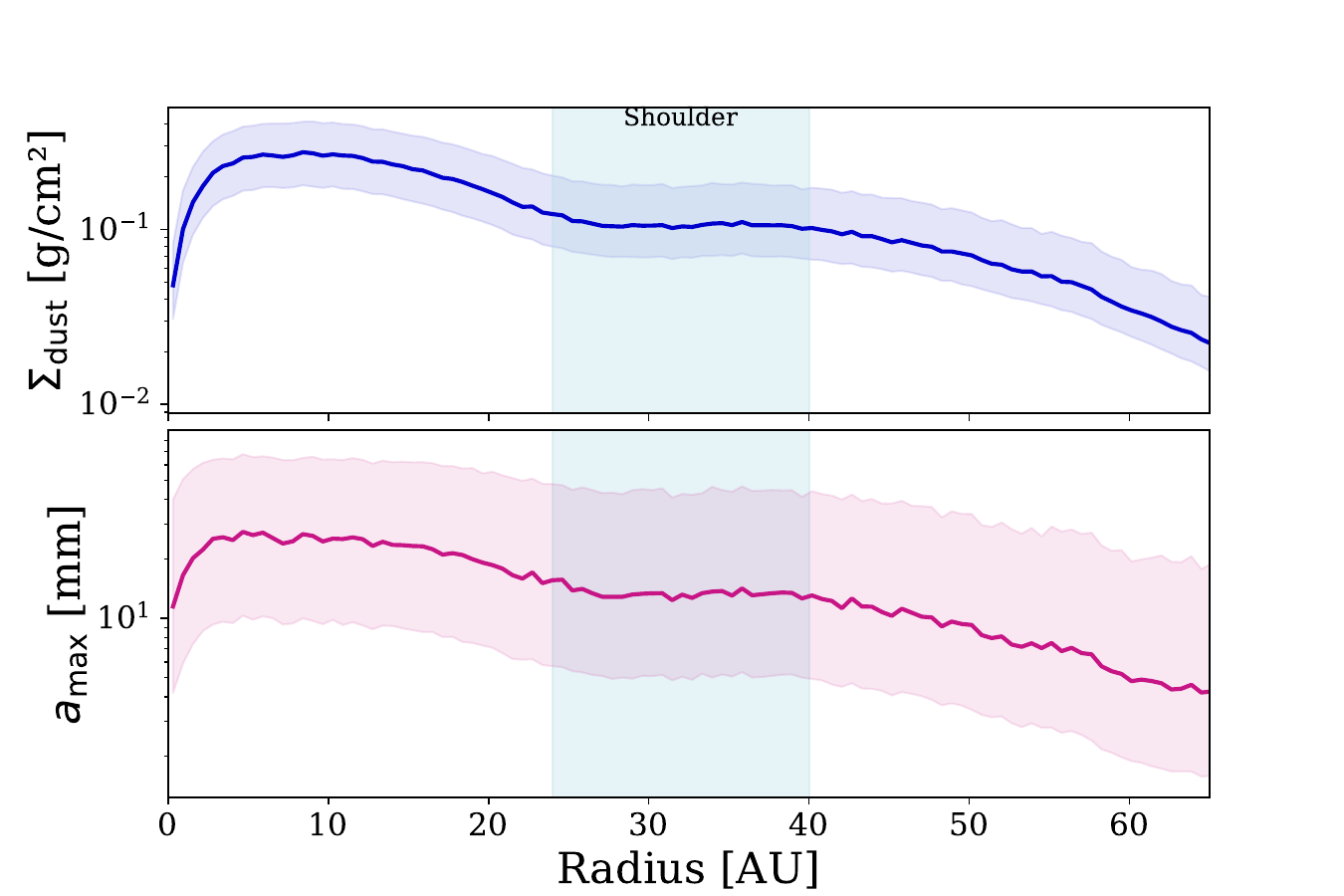}
    \caption{CY Tau}
    \label{fig:properties_CYTau}
    \end{subfigure}
    \caption{Dust properties of DoAr 25 (top) and CY Tau (bottom): the surface density in blue and the maximum-emitting grain size  in pink, derived from the dust continuum modeling. The temperature prior is a set power law and the maximum emitting grain size range is set to 1mm-10cm. The shoulder feature is noted on the plot with a light blue color.}
    \label{fig:properties_combined}
\end{figure}

\begin{figure}[htbp!]
    \centering
    \begin{subfigure}{\linewidth}
        \centering
    \includegraphics[width=\linewidth]{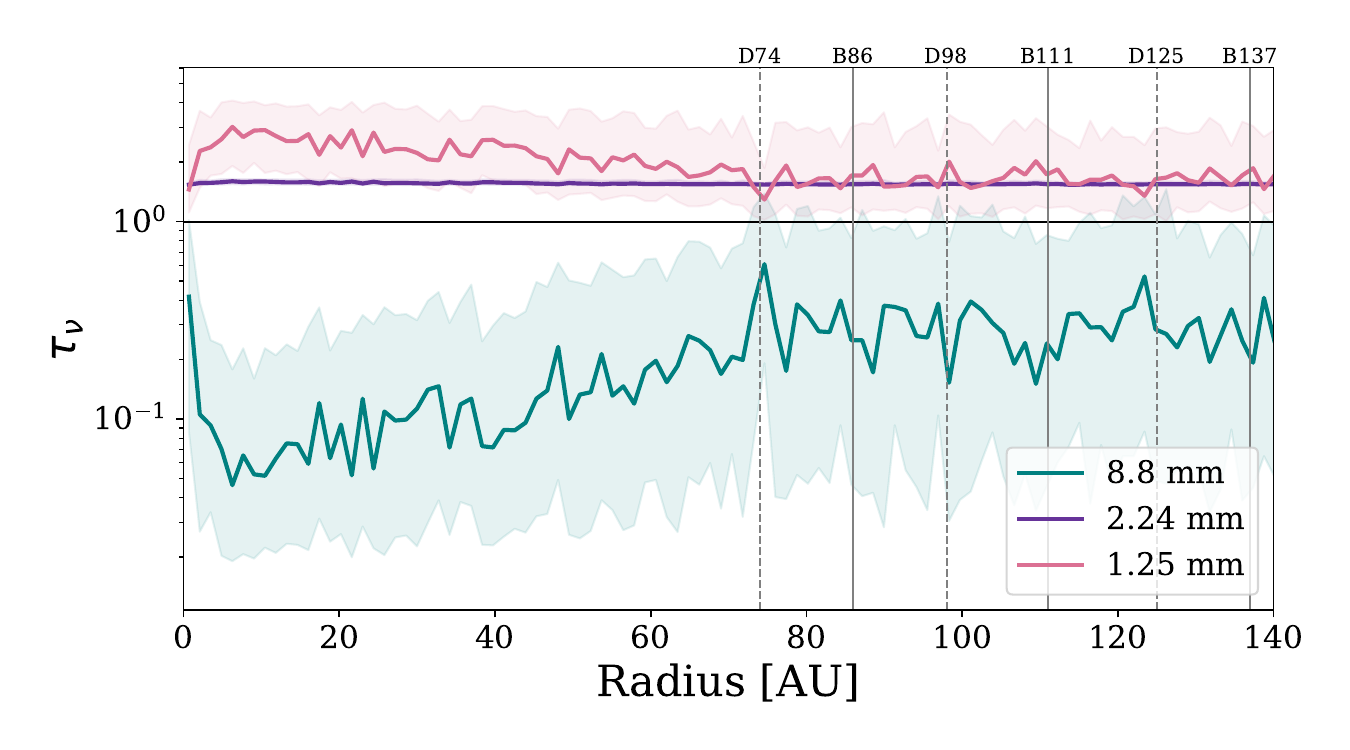}
    \caption{DoAr25}
    \label{fig:tau_large_doar25}
     \end{subfigure}

    \begin{subfigure}{\linewidth}
    \centering
    \includegraphics[width=\linewidth]{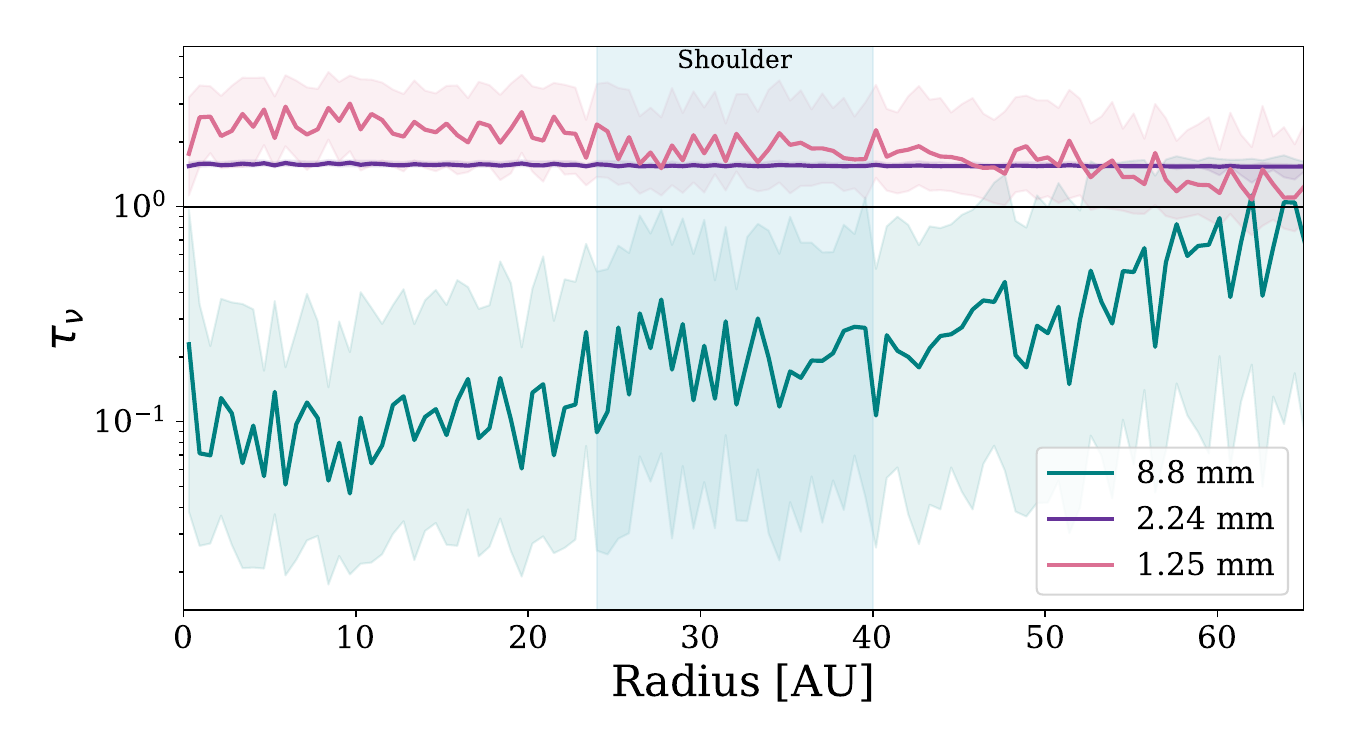}
    \caption{CY Tau}
    \label{fig:tau_large_cytau}
    \end{subfigure}
    \caption{Optical depth of DoAr 25 (top) and CY Tau (bottom) across multiple wavelengths, Band 4 in pink, Band 6 in teal, and VLA in purple. On the plot a straight line is drawn for $\tau=1$ to distinguish between optically thick and optically thin emission. With this we distinguish that the VLA is optically thin and the ALMA bands are optically thick. The shoulder feature is again noted with a light blue color. The temperature prior is a set power law and the maximum emitting grain size range is set to 1mm-10cm.}
    \label{fig:tau_large_combined}
\end{figure}

With a fixed temperature prior and fitting within the large grain size range, Fig.~\ref{fig:properties_combined} show smoother distributions. Particularly, Fig.~\ref{fig:properties_doar25} exhibits a bump coinciding with DoAr 25's main ring at 111 AU, while Fig.~\ref{fig:properties_CYTau} shows a slight dip corresponding to CY Tau's shoulder feature. The same behavior is observed in the surface density profiles shown in Figs.~\ref{fig:properties_doar25_small} and \ref{fig:properties_CYTau_small}, with more constrained uncertainties in the dust surface density profiles. However, the maximum emitting grain size cannot be reliably constrained when modeling in this smaller grain size range. Examining the optical depth plots (Fig.~\ref{fig:tau_large_combined}), 
VLA emission is optically thin while emission from ALMA bands is optically thick, as expected. Beyond this main difference, no other distinguishing features are apparent in the optical depth distributions. From here on out when referring to the results of DoAr 25 and CY Tau, we refer to the results derived from our second approach of radiative transfer modeling, using a power-law temperature prior and fitting the data within the large grain size regime.

\section{Discussion}\label{sec:discussion}

\subsection{The disk around DoAr 25}

The millimeter emission of DoAr~25 observed with ALMA is more extended than that seen in the VLA data-at 8.8mm-by 2" along the disk major axis. Comparing also with the CARMA and SMA observations (at 2.8 and 0.9mm respectively) presented in \cite{perez2015grain} the radial extent of the emission is comparable to what we see in the new ALMA observations at 1.3 and 2.2mm. No structure is reported in that work due to insufficient angular resolution to resolve substructures, given the synthesized beam, combined with the lower sensitivity of CARMA and the SMA relative to ALMA. The beam sizes are $0.48'' \times 0.35''$ (SMA-0.9mm), $0.64'' \times 0.33''$ (CARMA-2.8mm), and $0.038'' \times 0.033''$ (ALMA-1.3, 2.2mm), with corresponding sensitivities of 3.5\,mJy\,beam$^{-1}$, 0.27\,mJy\,beam$^{-1}$, and $\sim 10^{-6}$\,Jy\,beam$^{-1}$, respectively. The reason we do not see the extended emission in the VLA observations is likely not due to the radial drift of the dust particles, but due to the emission being optically thick at millimeter wavelengths, considering the optical depth plots shown in Figure~\ref{fig:tau_large_doar25}. This Figure clearly indicates that the VLA emission is optically thin whereas the emission from the ALMA bands is optically thick. When extracting the radial profiles from the image one prominent ring emerges at 111 AU that was also reported in \cite{huang2018disk}. The presence of a ring at similar radii across multiple wavelengths suggests that grains of different sizes are responding to the same underlying pressure structure. This consistency across wavelengths supports the interpretation that the observed morphology is shaped by substructure in the disk rather than by pure radial drift alone. 

We derive a spectral index of $\alpha\sim$1.8 at 1.25-2.4 mm throughout the disk and $\alpha$<2 at 1.25-8.8 mm and 2.4-8.8 mm in the inner region, <25 AU (Figure \ref{fig:spectral_index_doar25}). This value is usually associated with optically thick emission with high albedo, dominated by self-scattering \citep{ueda2020scattering}. Since this value at 1.25-2.4 mm is seen across the whole disk, this implies that the emission at millimeter wavelengths is optically thick everywhere, which is consistent with the optical depth plots. Alternatively, the low spectral index observed in the central $<25$ AU may be partly influenced by free-free emission from ionized gas close to the star, as suggested for several other disks \citep{macias2021characterizing, Guidi2022, Rota2024} and likely linked to to accretion-driven MHD wind emission from close to the star \citep{Rota2024, Rota2025}.

In the multiwavelength analysis (Fig.~\ref{fig:properties_doar25}), we get constrained results when we only leave the surface density and maximum emitting grain size as the free parameters, while setting the temperature profile to be a power law described by Eq.~\ref{eq:power_law}, with the grain size range being 1mm-10cm. The surface density in this case is rather flat with a notable bump at the location of the major dust ring at 111 AU. This bump is also seen in the ${\rm a}_{\text{max}}$ profile, however only tentatively since it lies within the error bars. If this result reflects an actual disk feature and is not just a statistical artifact, we can say that it is consistent with signatures predicted by dust trapping mechanisms \citep{2012A&A...538A.114P, sierra2025high}.

These results contradict the previous interpretation of the dust evolution mechanisms in DoAr~25. \citet{perez2015grain} concluded that DoAr 25 is dominated by radial drift as larger grains are concentrated in the inner disk and smaller grains dominate the outer disk, but with our radiative transfer modeling results we see that cm-sized grains dominate all the way to the outer disk. This is consistent with the existence of a dust trap located at 111 AU that halts the radial drift of grains toward the central star.

\subsection{The disk around CY Tau}

In CY Tau we observe the same phenomenon as in DoAr 25, where the dust emission is more extended in the ALMA image than in the VLA image. From Fig.~\ref{fig:tau_large_cytau} we can see that the emission in the VLA wavelength range is optically thin, whereas in the ALMA bands is optically thick. CY Tau is a lot more compact compared to DoAr 25, as its extent is $\sim$1" across the whole disk, whereas the extent of DoAr 25 is $\sim$2". The extent of the emission is comparable to the CARMA-1.3 and 2.8mm observations presented in \cite{perez2015grain}, but the beam sizes are much larger than those of the new ALMA data ($0.29'' \times 0.24''$ at 1.3\,mm and $0.40'' \times 0.35''$ at 2.8\,mm, compared to $0.06'' \times 0.041''$ at 2.2\,mm and $0.079'' \times 0.041''$ at 1.25\,mm). This resolution allows us to tentatively distinguish a gap in the Band 4 data. This gap  presents itself when extracting the radial brightness profiles as a bump or a shoulder feature, spanning between 20-40 AU. This structure is not seen in the Band 6 or VLA data. The lack of a clearly resolved shoulder in the other images does not necessarily imply the absence of pressure traps in CY Tau. Given the compact nature of the disk and the optical depth  at ALMA wavelengths, underlying substructures may be partially hidden, making the disk appear more drift dominated than it intrinsically is. Aside from this tentative gap, the rest of the disk is smooth and flat. We also see that the emission in Band 6 is slightly more extended than the emission in Band 4 indicating that there is some radial drift of larger grains toward the center.

For CY Tau, we derive a spectral index of $\alpha < 2$ across all wavelength combinations out to $\sim$50 AU (Fig.~\ref{fig:spectral_index_cytau}), followed by a sharp increase at larger radii. As in the DoAr 25 case, this value is a signature of optically thick emission with high albedo, likely dominated by scattering processes \citep{ueda2020scattering} or it may also be influenced by free-free emission associated with ionized gas or an accretion-driven wind \citep{macias2021characterizing, Guidi2022, Rota2024, Rota2025}. The rise in $\alpha$ beyond $\sim$50 AU suggests a transition to optically thin emission, possibly due to a drop in surface density and/or a change in grain size distribution near the edge of the disk. This behavior is consistent with a scenario where large grains are concentrated toward the inner regions due to radial drift, in contrast to DoAr~25.

Considering the multiwavelength modeling results, we see that both the dust surface density and maximum emitting grain size decrease monotonically with the radius. This is again consistent with larger grains drifting inward as the outer disk seems to be dominated by particles smaller than 1 cm. The only difference we can observe in these radial profiles is a slight enhancement at the shoulder region. This result lies within the error bars but if it reflects an actual disk feature it could indicate the presence of gap or cavity causing the accumulation of material at its edge. This feature is likely obscured in the Band 6 data because emission is more optically thick so it remains undetectable at that wavelength.

\subsection{Comparative analysis}

In this section, we present our results in the broader context of substructures in protoplanetary disks. We compare our findings to a broader sample of transition and full disks for which high-resolution multiwavelength analysis has been performed, from \citet{jiang2024grain} and \citet{2024ApJ...974..306S}. With "full disks," we refer to those often described as ringed or gapped -- disks that lack large inner cavities like transition disks but still exhibit narrow rings and gaps in their structure. The full disks included in \citet{jiang2024grain} have been previously studied through multiwavelength analyses, which constrained the radial distribution of dust emission and grain growth using spatially resolved continuum observations at millimeter wavelengths (e.g., \citealt{perez2015grain, trotta2013constraints}). Building on these works, \citet{jiang2024grain} reanalyzed the available multiwavelength data using a more comprehensive modeling framework, combining observations across multiple bands to better constrain the radial variations of grain size and surface density, and to reduce degeneracies present in earlier studies. The results used both in that work and in this study are taken from \citet{teague2022mapping} for TW Hya, \citet{pinte2015dust} for HL Tau, and from \citet{oberg2021molecules, teague2021molecules} for the five MAPS sources (GM Aur, IM Lup, AS 209, HD 163296, and MWC 480). In Table~\ref{tab:disk_summary} we summarize the radial substructures observed in all disks from the comparison sample, along with the proposed explanations for their properties. Our own  interpretations for DoAr 25 and CY Tau are also included.

\begin{table*}[htbp!]
\centering
\caption{Summary of the substructures observed in the studied sample of full and transition disks, and the explanations proposed in the literature for their origin.}
\begin{tabular}{lccc}
\hline\hline
\textbf{Disk Name} & \textbf{Substructures}\tablefootmark{a} & \textbf{Possible Origin} & \textbf{References}\tablefootmark{b} \\
\hline

\multicolumn{4}{c}{\textbf{Literature -- Transition disks}} \\
\specialrule{1.5pt}{0pt}{0pt}

CQ Tau & Gas cavity (50 AU) & Grain growth; Dust trap companion/dead zone & 1,2 \\

DM Tau & Ring; Gas cavity (25 AU) & Grain growth; Dust trap companion/dead zone & 1,3 \\

GM Aur & Large cavity (40 AU) & Grain growth; Dust trap companion/dead zone & 4,9 \\

LkCa 15 & Gas cavity (76 AU); Multiple rings & Companion; Dust traps; Dead zones & 1,5 \\

RXJ 1615 & Gas cavity (30 AU); Multiple rings & Dust trap companion/dead zones & 1,6 \\

SR24S & Gas cavity (35 AU); Ring & Grain growth; Dust trap companion/dead zone & 1,7 \\

UX Tau A & Gas cavity (31 AU); Ring & Dust trap companion/dead zone & 1 \\

\hline
\multicolumn{4}{c}{\textbf{Literature -- Full disks}} \\
\specialrule{1.5pt}{0pt}{0pt}

AS 209 & Multiple rings & Snow lines; Potential planets & 4,10,11,9 \\

HD 163296 & Multiple rings & Planet-disk interaction & 4,12,11,9 \\

HL Tau & Multiple rings & Planet-disk interaction & 13,11,9 \\

IM Lup & Wide ring & Pressure bump? & 4,12,11,9 \\

MWC 480 & Ring + gap & Tentative planet-disk interaction & 4,14,9 \\

TW Hya & Central cavity; Multiple rings & Accumulation of mm/cm particles; Pressure bumps & 8,9 \\

\hline
\multicolumn{4}{c}{\textbf{This Work}} \\
\specialrule{1.5pt}{0pt}{0pt}

CY Tau & Shoulder feature & Smooth disk & -- \\

DoAr 25 & Multiple rings & Dust trap & -- \\

\hline
\end{tabular}

\caption*{%
\\ \textbf{Notes.} \tablefoottext{a}{For the cavities of the transition disks, their size is indicated in the parenthesis.} \tablefoottext{b}{When multiple references are listed, the first typically corresponds to the detection of the disk substructure (e.g., cavity or rings), while subsequent references discuss the possible physical interpretation (e.g., grain growth, planet-disk interaction, or snow lines). \\

\textbf{References.} 
(1) \cite{2024ApJ...974..306S}; 
(2) \cite{ubeira2019dust}; 
(3) \cite{francis2022gap}; 
(4) \cite{Zhang2015} (MAPS); 
(5) \cite{leemker2022gas}; 
(6) \cite{van2015gas}; 
(7) \cite{pinilla2016can}; 
(8) \cite{macias2021characterizing}; 
(9) \cite{jiang2024grain}; 
(10) \cite{guzman2018disk}; 
(11) \cite{huang2018disk}; 
(12) \cite{izquierdo2023disc}; 
(13) \cite{wang2020architecture}; 
(14) \cite{liu2019ring}.}}

\label{tab:disk_summary}
\end{table*}

\begin{figure}[!htbp]
    \centering
    \includegraphics[width=1\linewidth, height=0.85\linewidth]{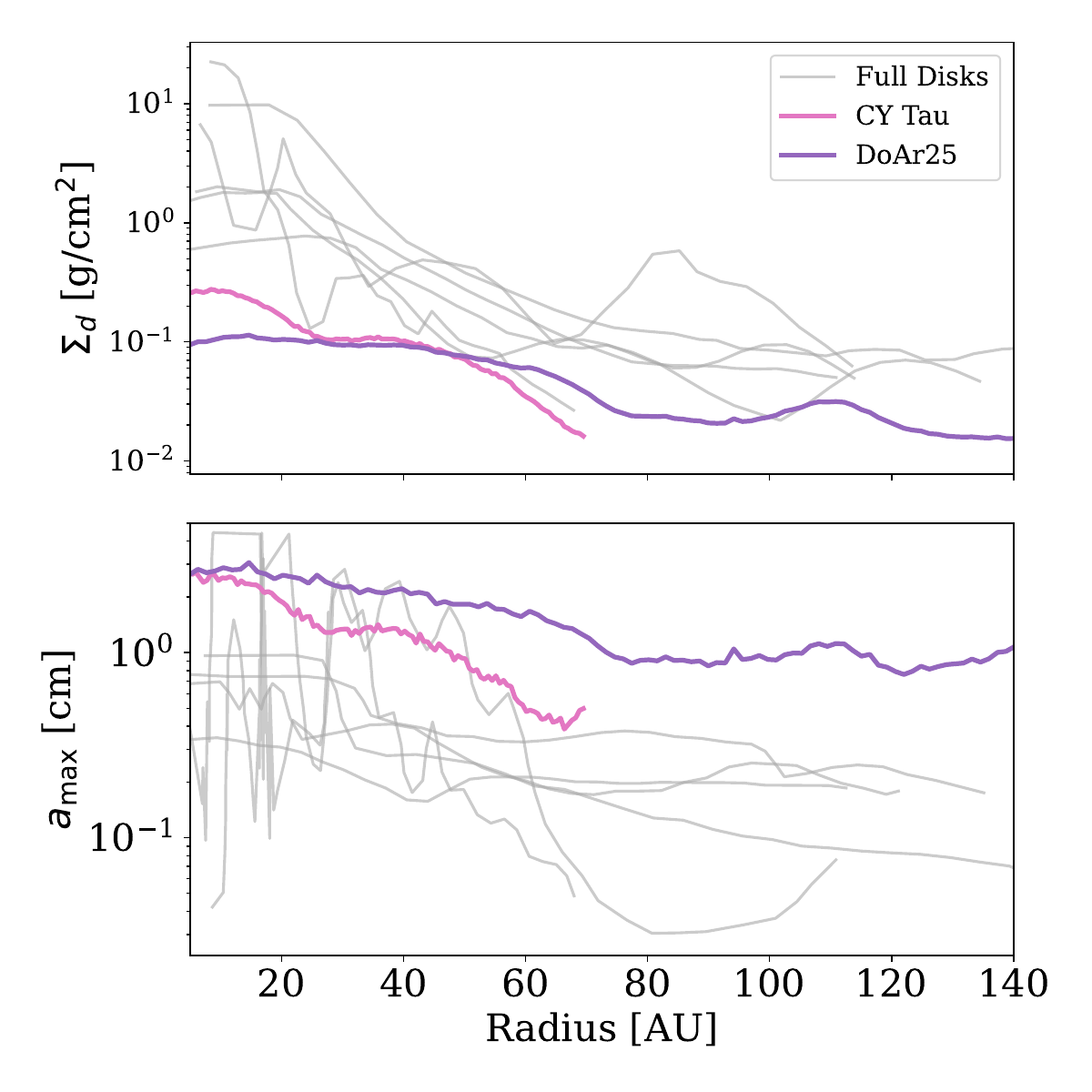}
    \caption{Comparison between the disks analyzed in this work, DoAr 25 (purple) and CY Tau (pink), and the sample of full disks (gray) from \citet{jiang2024grain}. The top panel shows the dust surface density profiles, highlighting that DoAr 25 and CY Tau exhibit significantly lower values compared to the full disk sample. The bottom panel displays the maximum grain size of the emitting dust, where both disks show notably larger grains than those in the comparison sample.}
    \label{fig:full_disks_comparison}
\end{figure}

\begin{figure}[!htbp]
    \centering
    \includegraphics[width=1\linewidth, height=0.85\linewidth]{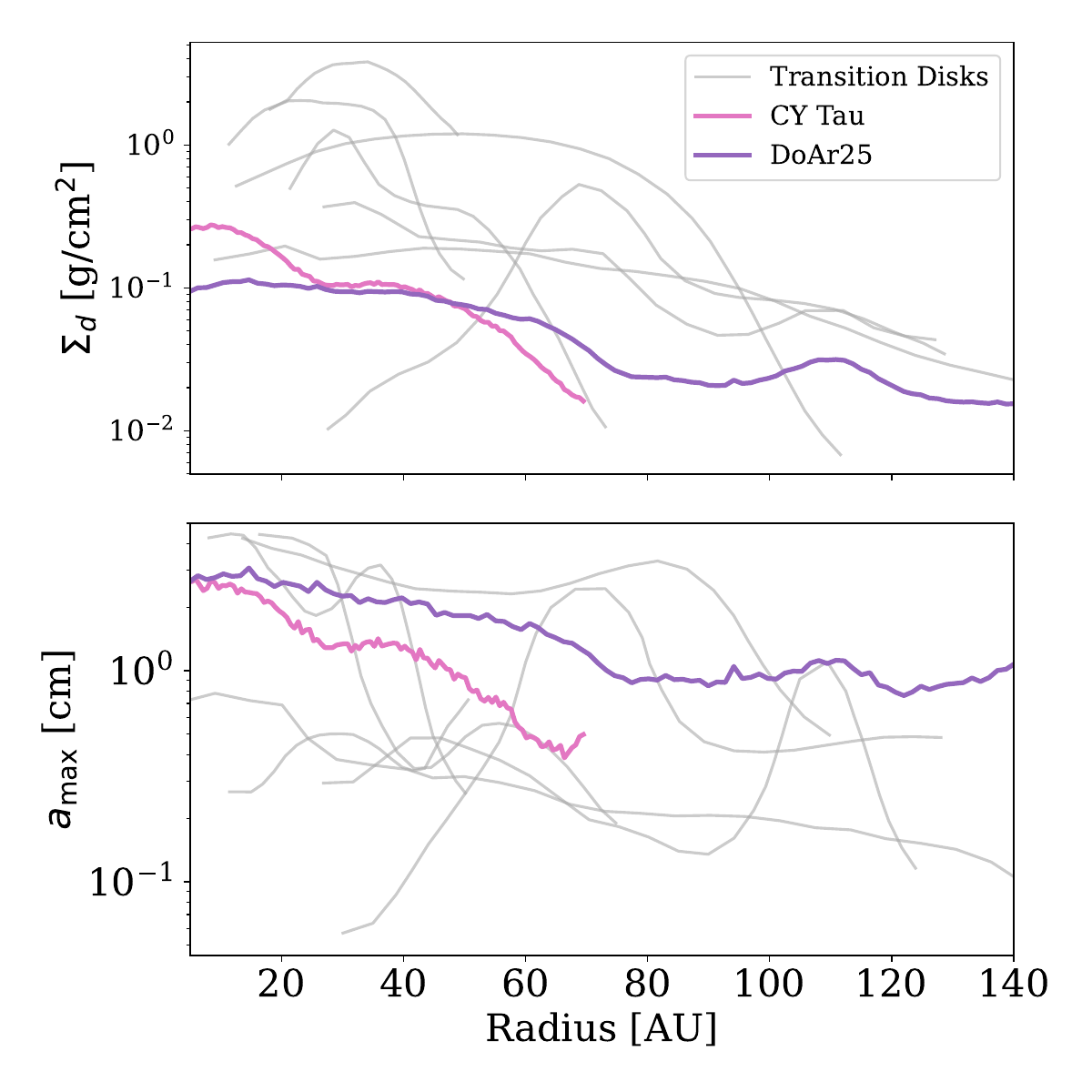}
    \caption{Comparison between the disks analyzed in this work, DoAr 25 (purple) and CY Tau (pink), and the sample of transition disks (gray) from \citet{2024ApJ...974..306S}. The grain properties are derived from the cavity radius outward, as given in Table~\ref{tab:disk_summary}. The top panel shows the dust surface density profiles, highlighting that DoAr 25 and CY Tau exhibit lower values compared to the transition disk sample. The bottom panel displays the maximum grain size of the emitting dust, where both disks show larger grains than most transition disks.}
    \label{fig:transition_disks_comparison}
\end{figure}

\begin{figure}[htbp!]
    \centering
    \includegraphics[width=0.9\linewidth, height=0.7\linewidth]{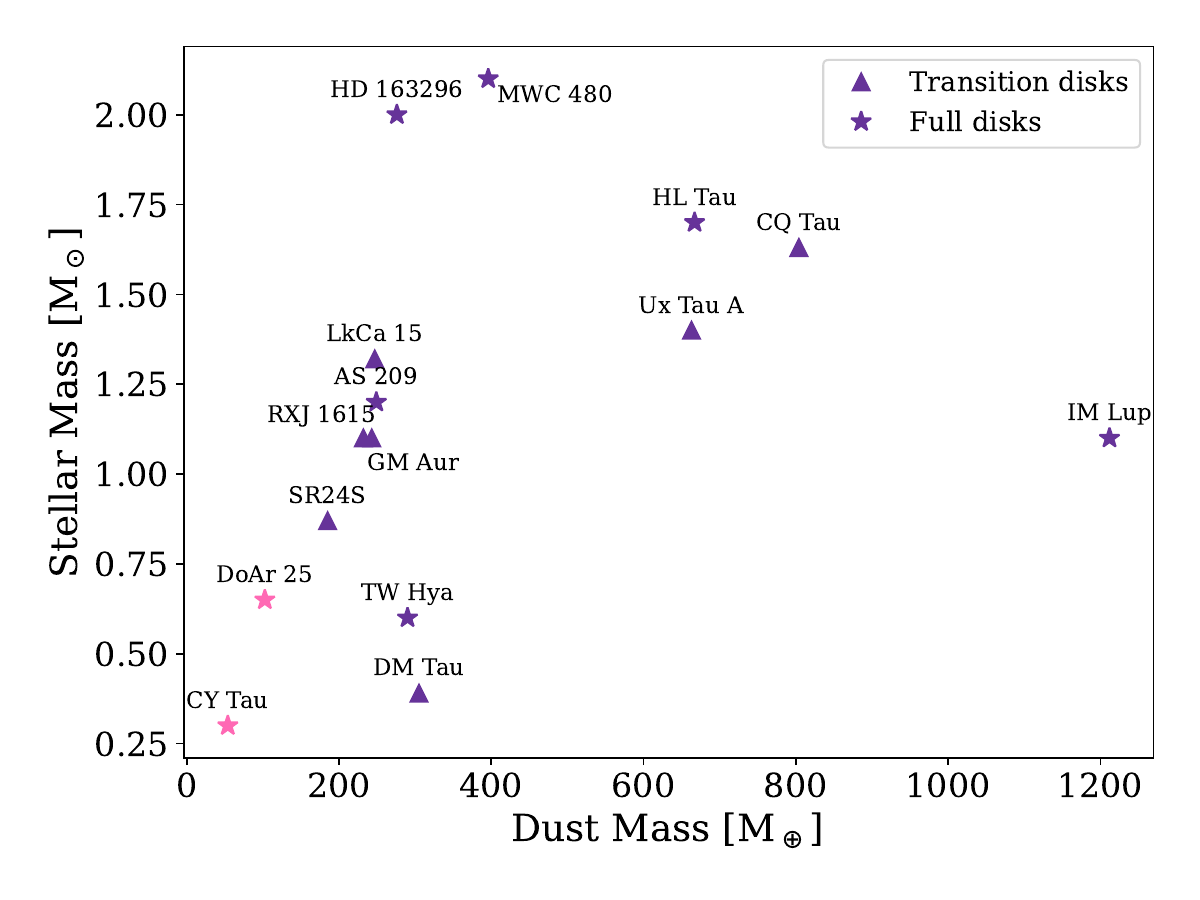}
    \caption{Dust masses of the disks in the sample %in Earth masses 
    plotted against the central stellar mass. %in solar masses. 
    Transition disks are plotted with triangles and full disks with stars. CY Tau and DoAr 25 are highlighted with the pink color. For references see Table~\ref{tab:disk_summary} and \citet{2024ApJ...974..306S, guerra2024into}. The values used for the plot can be found in Table \ref{tab:dust_masses}.}
    \label{fig:dust_mass_star_mass}
\end{figure}

DoAr~25 and CY Tau exhibit lower dust surface densities compared to the disks from \citet{jiang2024grain} and \citet{2024ApJ...974..306S}, as shown in Figures~\ref{fig:full_disks_comparison} and \ref{fig:transition_disks_comparison}. The dust  masses in  Fig.~\ref{fig:dust_mass_star_mass}, computed with multiwavelength analysis, support this, showing that DoAr 25 and especially CY Tau have lower total dust masses and significantly lower stellar masses compared to previously studied targets. This suggests that the reduced dust surface density can likely be attributed to a simple lack of material; with less dust available, the disks naturally exhibit lower surface densities.


A positive correlation between total dust mass and stellar mass has been established in several studies \citep[see e.g., Fig.~4 in][]{manara2022demographics}. Disks around lower-mass stars tend to have lower total (dust) masses, as shown in ALMA surveys across multiple star-forming regions. However, the significant spread seen at similar stellar mass may be due to an intrinsic scatter in initial disk masses or may reflect evolutionary effects such as variations in efficient grain growth and radial drift, both of which can reduce the observable dust mass \citep{pinilla2020hints, pinilla2025observational}.

As shown in Fig.~\ref{fig:full_disks_comparison} (separate profiles are shown individually in Fig.~\ref{fig:regular-sigma}), most full disks in the sample have surface densities in the range of $0.1$-$1\,\text{g/cm}^2$ and the profile reduces monotonically with radius often with a steep drop near the center, with HL Tau, IM Lup and TW Hya standing out as clear exceptions: all disks reach surface densities up to $\sim 10\,\text{g\,cm}^{-2}$, HL Tau shows a second enhancement at $\sim 85$\,AU, while TW Hya exhibits a steep decline out to $\sim 25$\,AU. Both DoAr 25 and CY Tau exhibit surface densities nearly an order of magnitude lower, in the range of $0.01$--$0.1\,\text{g\,cm}^{-2}$, while maintaining the same overall monotonically decreasing radial profile. In some cases, such as HL Tau and HD 163296, strong bumps are observed in the radial profiles. These features are similar to those seen in DoAr 25. 

Looking at the maximum grain size in Fig.~\ref{fig:full_disks_comparison} (with the individual profiles shown in Fig.~\ref{fig:regular-alpha}), the majority of full disks show $a_{\text{max}}$ values between $0.1$ and $1\,\text{cm}$. Our two disks lie toward the upper end of this distribution, with grain sizes reaching values approximately an order of magnitude above the typical range for most disks in the sample. However, they are not unique in this respect, as TW Hya and HL Tau also reach comparable values. TW Hya is likewise a low-mass star hosting a relatively low-mass disk, while HL Tau is a much younger system, which may help explain its similarly large grain sizes. Additionally, the radial profiles of the comparison sample are generally smooth. This behavior is consistent with their low dust surface densities: if a significant fraction of the dust mass is concentrated in larger grains, less small-grain material remains to contribute to the surface density measured by continuum emission. This suggests that despite their lower total dust content, grain growth has proceeded efficiently in these disks, potentially aided by the dust trap at 111 AU in DoAr 25 and the potential presence of a gap causing the shoulder feature in CY Tau. Again, TW Hya stands out for its pronounced radial variations in $a_{\text{max}}$. This comparison may be subject to bias, as the aforementioned studies do not incorporate longer-wavelength data as we do here, which in turn affects the predicted maximum grain size, longer wavelengths probe larger grains \citep[e.g.,][for HL Tau]{carrasco2019radial}. However, \citet{ueda2025multiwavelength} find for the same disk that even with including longer wavelengths in their analysis a smaller grain interpretation is possible and it depends on the dust composition chosen \citep[see also][]{shi2026probing,zagaria2025multi,guidi2022distribution}. This shows that the choice of dust composition can impact the analysis just as much as the choice of wavelengths included.

Taken together, the combination of lower surface densities and significantly larger maximum grain sizes in DoAr 25 and CY Tau suggests that a substantial fraction of the dust mass may be locked in large grains that contribute less efficiently to the observed continuum emission. This behavior is consistent with dust evolution models in which disks around lower-mass stars are more strongly affected by radial drift, leading to rapid inward transport of millimeter-sized grains in the absence of efficient trapping \citep{2020A&A...635A.105P}. In this framework, the smooth morphology and compact size of CY Tau suggest that its dust evolution is dominated by radial drift, limiting the retention of millimeter-sized particles. In contrast, the prominent substructures observed in DoAr 25 are consistent with the presence of pressure maxima that can act as efficient dust traps, enabling grain growth to sizes beyond those probed at millimeter wavelengths and reducing the observable surface density despite the presence of significant solid mass. The differing behaviors of these two systems therefore provide evidence for distinct dust evolution regimes within lower-mass disks, where radial drift dominates in the absence of strong traps, while efficient grain growth can occur in disks hosting significant substructure.

This suggests that the observed diversity in dust properties among lower-mass disks may be driven not only by differences in total dust mass, but also by the relative efficiency of radial drift and dust trapping. In this context, disks that appear dust-poor at millimeter wavelengths may not necessarily lack solid material, but instead may have undergone efficient grain growth, shifting a significant fraction of the dust mass to larger, less observable sizes. These results highlight the critical role of substructure in enabling grain growth in low-mass disks and provide observational support for the theoretical predictions of dust evolution models \citep{2020A&A...635A.105P}.

We also compare our results to the sample of transition disks. The grain properties are derived from the cavity radius outward, as given in Table~\ref{tab:disk_summary}. In this case, the dust surface densities (shown separately in Fig.~\ref{fig:transition-sigma}) of DoAr 25 and CY Tau are more comparable with the sample of transition disks, although they still tend to be on the lower end. The surface density profiles for most transition disks show minimal radial variation outside of the cavity region. Most of these disks exhibit gas cavities, references listed in Table~\ref{tab:disk_summary}, that are thought to act as dust traps, likely formed through interactions with embedded companions \citep{2012A&A...538A.114P,van2016resolved}.

For the maximum grain size (shown separately in Fig.~\ref{fig:transition-alpha}), our findings fall within the typical range observed in the transition disk sample. Many transition disks show a clear enhancement in the outer disk regions, consistent with the accumulation of large grains at cavity edges. Several disks also exhibit reduced grain sizes within the inner cavity, further supporting the interpretation that radial drift removes large particles from the inner regions while pressure trapping concentrates them at the cavity rim.

\subsection{Maximum grain size compared to growth barriers}

Dust evolution models predict that the growth of particles can be limited either by the fragmentation of particles due to turbulent velocities ($a_{\rm{frag}}$); or due to radial drift ($a_{\rm{drift}}$), where particles either experience high collisions from drift velocities or move toward the star in short time scales of the disk lifetime. These limits are given by \cite{birnstiel2010+models}, as

\begin{equation}
        a_{\mathrm{frag}}=\frac{2}{3\pi}\frac{\Sigma_g}{\rho_s \alpha}\frac{v_{\rm{frag}}^2}{c_s^2}.
  \label{eq:afrag}
\end{equation}
\noindent and
\begin{equation}
        a_{\mathrm{drift}}=\frac{2 \Sigma_d}{\pi\rho_s}\frac{v_K^2}{c_s^2}\left \vert \frac{\mathrm{d} \ln P}{\mathrm{d} \ln r} \right \vert^{-1}.
  \label{eq:adrift}
\end{equation}
respectively.

In Eq.~\ref{eq:afrag} $\Sigma_g$, $\rho_s$, $\alpha$, $c_s$, and $v_{\rm{frag}}$ correspond to the gas surface density, dust volume density, which for the DSHARP dust opacities assumed in this work is $\rho_s=1.6\,$g\,cm$^{-3}$, viscosity, sound speed, and fragmentation velocity, respectively. The fragmentation velocity is the speed in which particles are expected to fragment after collision, which depends on the composition of the grains. Currently, the values of $v_{\rm{frag}}$ from laboratory experiments and numerical simulations range from $\sim$1m\,s$^{-1}$ to $\sim$10m\,s$^{-1}$ \citep[e.g.,][]{blum2008}. The values of $\Sigma_g$ and $\alpha$ remain highly unconstrained from observations, and only available for a handful of disks \citep[e.g.,][]{rosotti2023, zhang2025alma}. On the other hand, for the drift barrier (Eq.~\ref{eq:adrift})  $\Sigma_d$ is the dust surface density, $\mathrm{d}\ln P/ \mathrm{d}\ln r$ is the pressure gradient, and $v_K$ the Keplerian velocity.

To compare the obtained maximum grain size from observations to these two growth barriers, we assume different  values for $v_{\rm{frag}}$, $\alpha$ and gas-to-dust mass ratio ($\epsilon$). The gas surface density is then assumed to be $\Sigma_g=\epsilon\Sigma_d$, and we used the $\Sigma_d$ that is inferred from the radiative transfer calculations. The gas disk pressure is calculated at the disk midplane as

\begin{equation}
    P=\rho c_s^2 \quad{\textrm{with}} \quad \rho=\frac{\Sigma_g}{\sqrt{2\pi}H_g},
  \label{eq:Pgas}
\end{equation}
\noindent where $H_g$ is the gas pressure scale height given by $H_g=c_s/\Omega$, with $\Omega$ being the Keplerian frequency. 

\begin{figure*}[htbp!]
    \centering
    \includegraphics[width=1\linewidth]{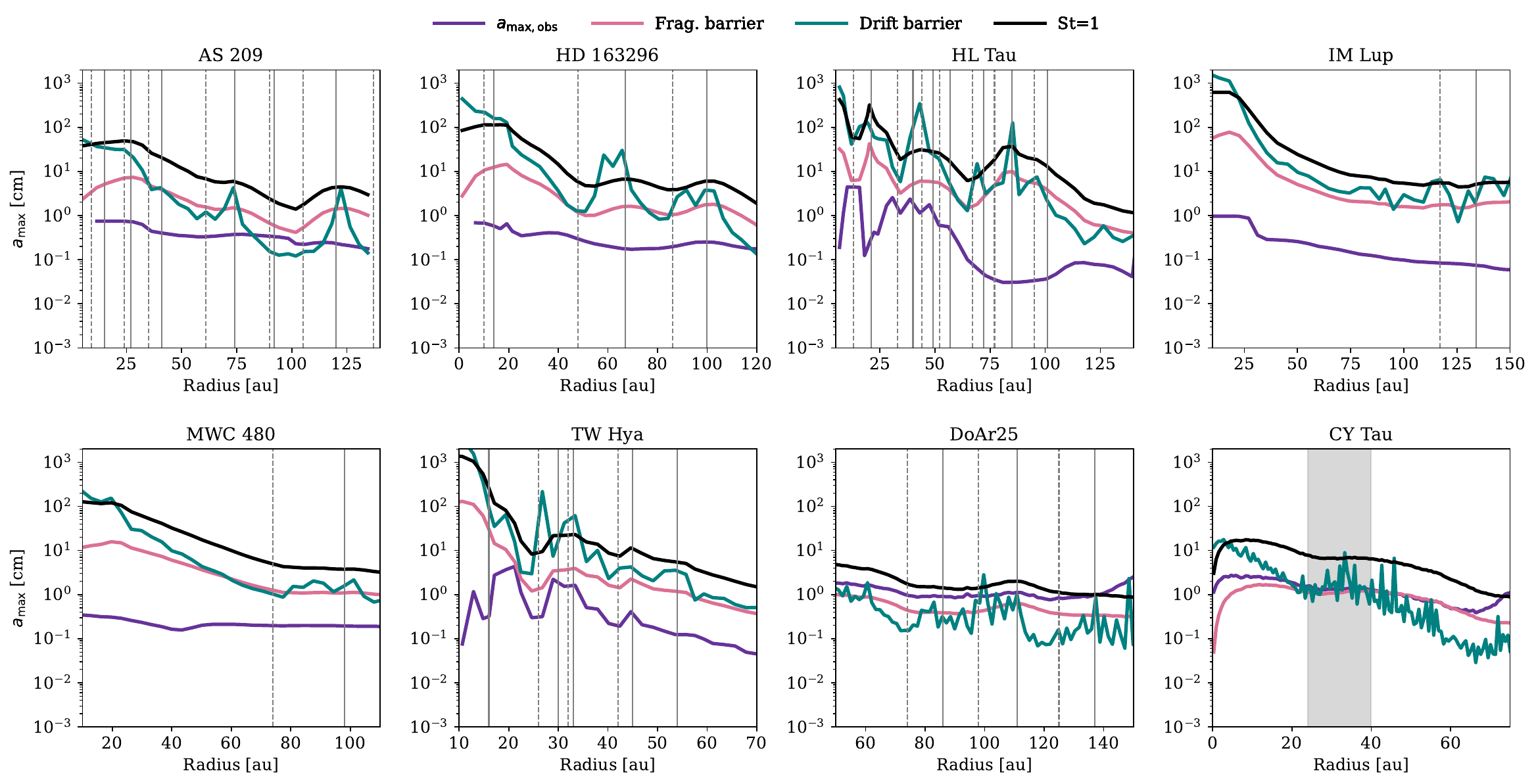}
    \caption{Radial profiles of the maximum grain size inferred from the radiative transfer modeling ($a_{\rm{max,obs}}$ in purple) compared with the theoretical fragmentation ($a_{\rm{frag}}$ in pink) and drift ($a_{\rm{drift}}$ is in teal) barriers for the sample disks. The barriers are calculated assuming $v_{\rm{frag}}=10$ m/s$^{-1}$, $\alpha=10^{-3}$ and g2d=100. The black line indicates the Stokes number limit St=1.}
    \label{fig:frag_barriers_v10}
\end{figure*}

Fig.~\ref{fig:frag_barriers_v10} shows the comparison of the maximum grain size from our radiative transfer models and the growth barriers ($a_{\rm{frag}}$ and $a_{\rm{drift}}$), assuming $v_{\rm{frag}}=10$m\,s$^{-1}$ and $\alpha=10^{-3}$, while Fig.~\ref{fig:frag_barriers_v1} shows the same with $v_{\rm{frag}}=1$m\,s$^{-1}$ and the same $\alpha$. In addition, the line corresponding to St=1 is also shown, where St is the Stokes number calculated at the midplane:

\begin{equation}
\textrm{St}=  \frac{a\rho_s}{\Sigma_g}\frac{\pi}{2}.
\label{eq:stokes}
\end{equation}

From these figures, it is clear that for most of the disks studied in \citet{jiang2024grain}, $a_{\rm{max}}$ is closer to the fragmentation limit ($a_{\rm{frag}}$) than to the drift limit ($a_{\rm{drift}}$) when $v_{\rm{frag}}=10$m\,s$^{-1}$,  suggesting that growth is limited by fragmentation by turbulent velocities. However, this is not the case for DoAr~25 and CY Tau, where $a_{\rm{max}}$ is  closer to $a_{\rm{drift}}$, suggesting that the maximum grain size in these two disks is limited by drift and supporting the idea that drift is shaping the evolution of the dust in these disks. If the fragmentation velocity is assumed to be $v_{\rm{frag}}=1$m\,s$^{-1}$, then $a_{\rm{frag}}$ is below $a_{\rm{max}}$, which lies between $a_{\rm{frag}}$ and $a_{\rm{drift}}$, except again for DoAr~25 and CY Tau, where $a_{\rm{max}}$ is much closer to $a_{\rm{drift}}$. Interestingly, the values of $a_{\rm{max}}$ for DoAr~25 and CY Tau are near St=1 (specially in comparison with the rest of the disks) implying that they are drifting close to the maximum possible drift velocities \citep{brauer2008coagulation}.  There are several assumptions in these calculations that remain unconstrained from observations ($\alpha$, $\epsilon$, $v_{\rm{frag}}$). This exercise is just a proof of concept that shows that when taking the same assumptions across all the disks, the obtained $a_{\rm{max}}$ in DoAr~25 and CY Tau agrees best with $a_{\rm{drift}}$, contrary to the other disks.

\section{Conclusions}\label{sec:cnclusions}
In this study, we investigated the physical properties of the protoplanetary disks around CY Tau and DoAr 25 using multiwavelength ALMA and VLA continuum observations from 1.3 to 8.8 mm. With multiwavelength dust continuum forward modeling, we assessed whether the dust evolution in these disks is primarily driven by radial drift or influenced by the presence of dust traps.

We constrained the radial distribution of two key physical parameters: the maximum grain size distribution and the dust surface density. We compared our findings with a broader sample of disks from the literature, for which multiwavelength analysis was already performed, including both full disks with ring-like substructures and transition disks. Our main conclusions are as follows:
\begin{enumerate}

\item The ALMA continuum emission is more extended than that of the VLA, which is likely due to optical depth effects and sensitivity limitations rather than radial drift.

\item A prominent dust ring is observed at 111 AU in DoAr 25, which is in agreement with results from the DSHARP survey \citep{andrews2020observations}. Radiative transfer modeling reveals a generally flat surface density profile, with a clear bump at 111 AU, consistent with the presence of a dust trap. The maximum grain size remains large (centimeter-sized) out to the outer regions of the disk, in contrast with earlier interpretations that suggested significant radial drift. These results support the presence of a dust trap at 111 AU, which slows inward drift and enables the accumulation of large grains.

\item The CY Tau system is more compact than DoAr 25, with ALMA emission more extended than  VLA emission, primarily due to optical depth and sensitivity differences. The disk presents a smooth and flat brightness profile with a shoulder feature between 20-40 AU, possibly indicative of a gap or marginal dust trap. Both the grain size and dust surface density decrease with radius, suggesting that CY Tau is largely dominated by radial drift, with limited retention of millimeter-sized particles in the outer disk.

\item When compared to a broader sample of full and transition disks, both DoAr 25 and CY Tau exhibit lower dust surface densities, likely reflecting their lower disk mass, consistent with their lower stellar mass.

\item Despite their lower surface densities, both disks host significantly larger maximum grain sizes than typically observed in similar disks. In DoAr 25, the prominent substructures act as dust traps, enabling efficient grain growth to centimeter-sized particles and retaining large grains in the outer disk. In contrast, CY Tau is fully dominated by radial drift, resulting in the inward migration of millimeter-sized grains and a more compact distribution. These findings are consistent with dust evolution models in which disks around lower-mass stars are more strongly affected by radial drift unless efficient dust traps are present.

\item Comparison with theoretical growth barriers suggests that grain growth in most disks from the sample is fragmentation-limited, whereas in DoAr 25 and CY Tau the inferred maximum grain sizes are more consistent with the drift barrier and approach St=1, indicating that radial drift is the dominant process regulating dust evolution in these disks.

\item Collectively, the differing behaviors of CY Tau and DoAr 25 provide evidence for distinct dust evolution regimes in lower-mass disks: radial drift dominates in the absence of strong substructures, whereas efficient grain growth can occur in disks hosting significant dust traps. This highlights that disks that appear dust-poor at millimeter wavelengths may still contain substantial solid mass in larger, less observable grains.
\end{enumerate}

Overall, this study provides valuable constraints and advances our understanding of the physical conditions in CY Tau and DoAr 25, showing that even low-mass disks can host large grains and dust traps. These results underline the diversity of disk evolution pathways and set the stage for future investigations targeting similar systems.

\begin{acknowledgements}
This paper makes use of the following ALMA data: ADS/JAO.ALMA\#2022.1.01284.S, ADS/JAO.ALMA\#2022.1.01302.S, ADS/JAO.ALMA\#2016.1.00484.L. ALMA is a partnership of ESO (representing its member states), NSF (USA) and NINS (Japan), together with NRC (Canada), NSTC and ASIAA (Taiwan), and KASI (Republic of Korea), in cooperation with the Republic of Chile. The Joint ALMA Observatory is operated by ESO, AUI/NRAO and NAOJ. A.T. is supported by a studentship from the Science and Technology Facilities Council (STFC) of the United Kingdom.
\end{acknowledgements}

\bibliographystyle{bibtex/aa}
\bibliography{bib}

\begin{appendix}
\section{Spectral Index}
\begin{figure}[htbp!]
    \centering
    \includegraphics[width=1\linewidth]{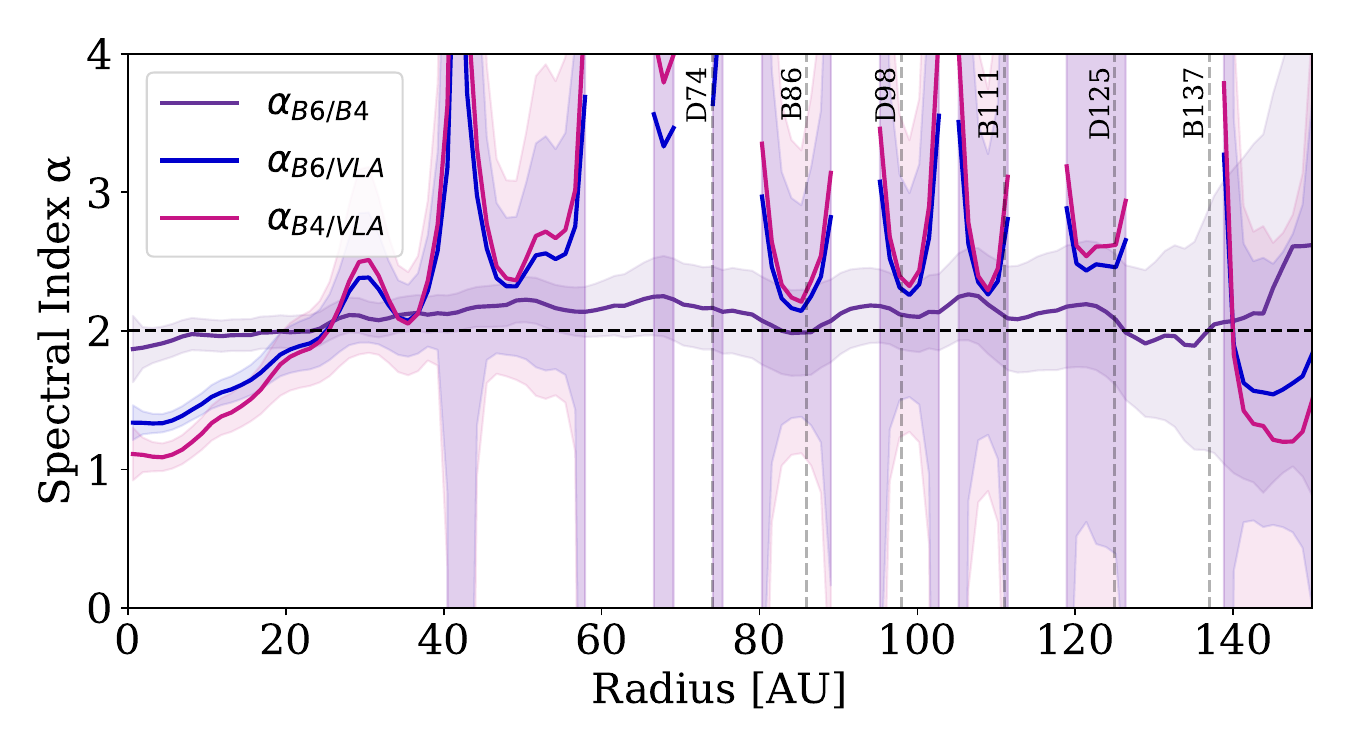}
    \caption{Spectral index radial profiles of the dust emission of the disk around DoAr 25 between 1.25 and 2.4 mm (purple), 1.25 and 8.8 mm (blue), and 2.4 and 8.8 mm (pink). These radial profiles were obtained following Equation \ref{eq:spectral_index}. A dashed black line has been drown at $\alpha=2$ to indicate the Rayleigh-Jeans limit.}
    \label{fig:spectral_index_doar25}
\end{figure}

\begin{figure}[htbp!]
    \centering
    \includegraphics[width=1\linewidth]{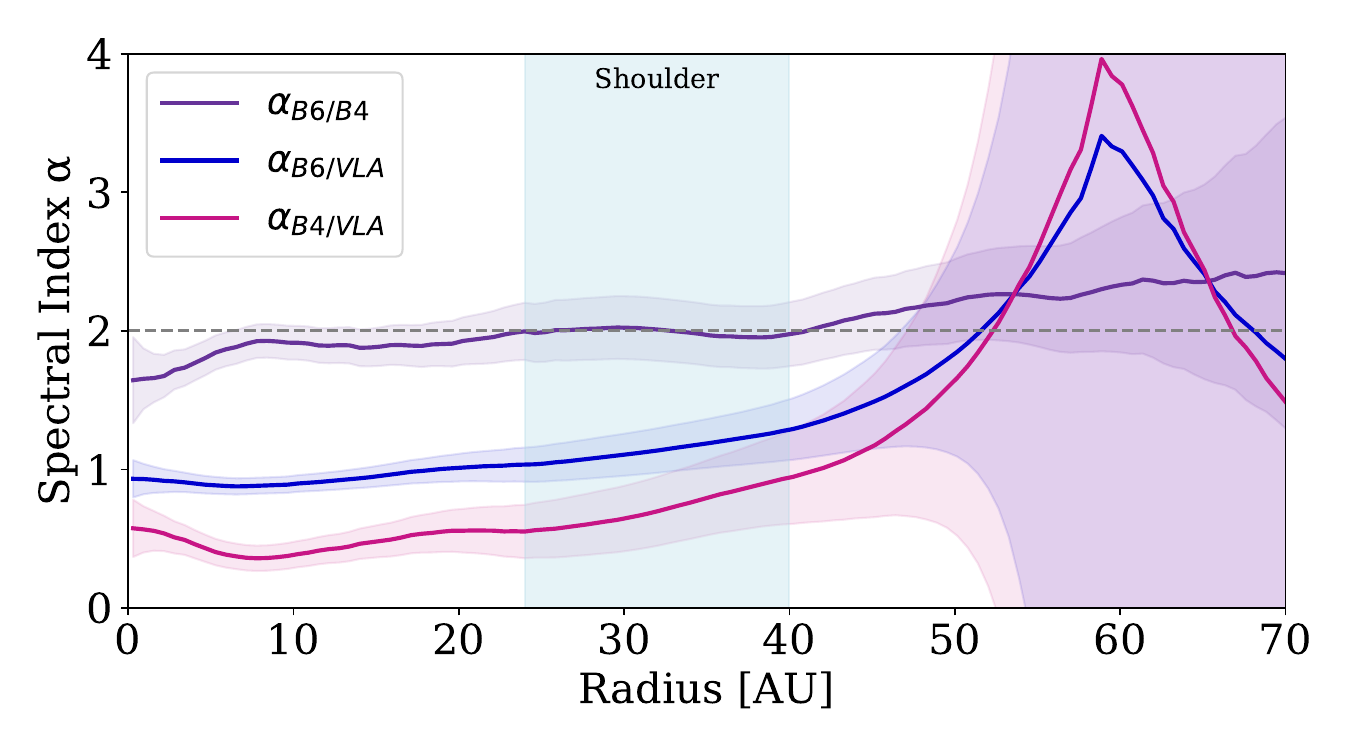}
    \caption{Spectral index radial profiles of the dust emission of the disk around CY Tau between 1.25 and 2.4 mm (purple), 1.25 and 8.8 mm (blue), and 2.4 and 8.8 mm (pink). These radial profiles were obtained following Equation \ref{eq:spectral_index}. The shoulder feature of the disk is indicated with light blue. The shaded color regions indicate the error of the mean at each radii. A dashed black line has been drown at $alpha=2$ to indicate the Rayleigh-Jeans limit.}
    \label{fig:spectral_index_cytau}
\end{figure}

We have calculated the spectral index $\alpha$ profiles, for both disks presented in Figs.~\ref{fig:spectral_index_doar25} and \ref{fig:spectral_index_cytau}, following Eq.~\ref{eq:spectral_index}

\begin{equation}\label{eq:spectral_index}
    \alpha = {\log _{10}}\,{{\left[ {{{{F_{{v_1}}}} \mathord{\left/ {\vphantom {{{F_{{v_1}}}} {{F_{{v_2}}}}}} \right. \kern-\nulldelimiterspace} {{F_{{v_2}}}}}} \right]} \mathord{\left/ {\vphantom {{\left[ {{{{F_{{v_1}}}} \mathord{\left/ {\vphantom {{{F_{{v_1}}}} {{F_{{v_2}}}}}} \right. \kern-\nulldelimiterspace} {{F_{{v_2}}}}}} \right]} {{{\log }_{10}}}}} \right. \kern-\nulldelimiterspace} {{{\log }_{10}}}}\left[ {{{{v_1}} \mathord{\left/ {\vphantom {{{v_1}} {{v_2}}}} \right. \kern-\nulldelimiterspace} {{v_2}}}} \right],
\end{equation}

where ${F_{{v_1}}}$ and ${F_{{v_2}}}$ are the flux densities measured in the observing wavelengths, $\nu_1$ and $\nu_2$, respectively. The spectral index is a helpful diagnostic when trying to disentangle the various physical origins of the emission \citep{testi2014protostars}.

We derive a spectral index of $\alpha\sim$1.8 at 1.25-2.4 mm throughout the disk and $\alpha<$2 at 1.25-8.8 mm and 2.4-8.8 mm in the inner region (<25 AU). This value is usually associated with optically thick emission with high albedo, dominated by self-scattering \citep{ueda2020scattering}. Since this value at 1.25-2.4 mm is seen across the whole disk shows that the emission coming from the ALMA bands is optically thick everywhere, which we also concluded earlier from the optical depth plots.

For CY Tau, we derive a spectral index of $\alpha < 2$ across all wavelength combinations out to $\sim$50 AU, followed by a sharp increase at larger radii. This low value is typically interpreted as a signature of optically thick emission with high albedo, likely dominated by scattering processes \citep{ueda2020scattering}. The rise in $\alpha$ beyond $\sim$50 AU suggests a transition to optically thin emission, possibly due to a drop in surface density and/or a change in grain size distribution near the edge of the disk. This behavior is consistent with a scenario where large grains are concentrated toward the inner regions due to radial drift.

\section{Supporting Figures and Data}

In this appendix, we present additional figures and data (Table~\ref{tab:dust_masses}) supporting the analysis. 

\begin{figure}[htbp!]

        \includegraphics[width=\linewidth]{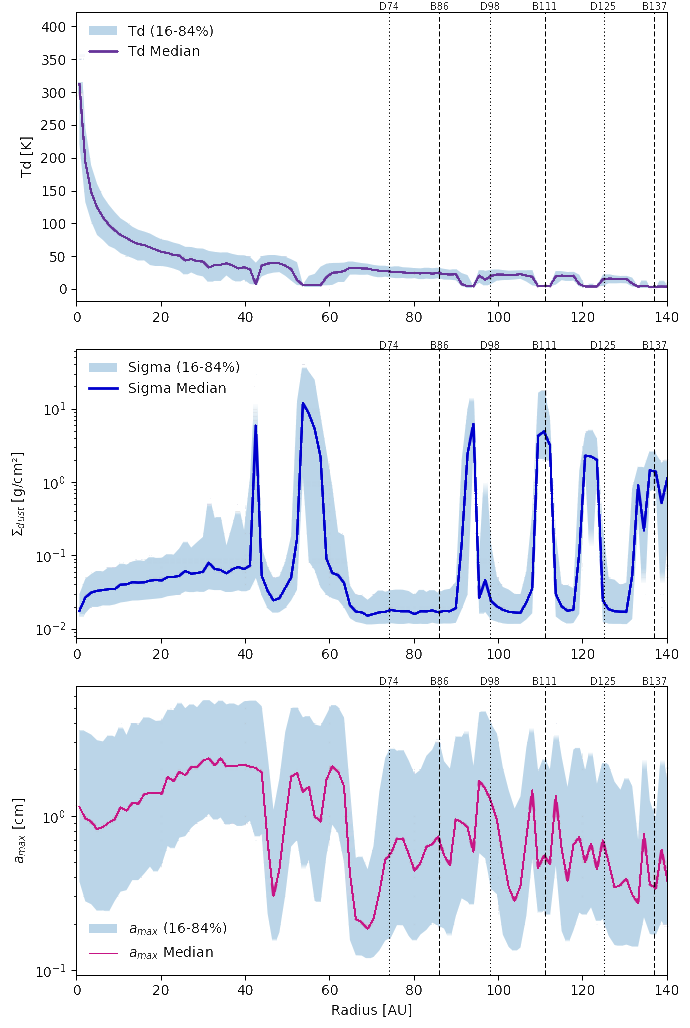}
        \caption{Physical properties ($T_d,\ \Sigma_{\text{dust}},\ a_{\text{max}}$) of DoAr 25 in $1mm-10cm$ grain sizes calculated with a physical temperature prior. The bright rings and gaps are again noted following the convention by \citep{huang2018disk}.}
        \label{fig:doar_large_prior}
\end{figure}

\begin{figure}[htbp!]
        \centering
        \includegraphics[width=\linewidth]{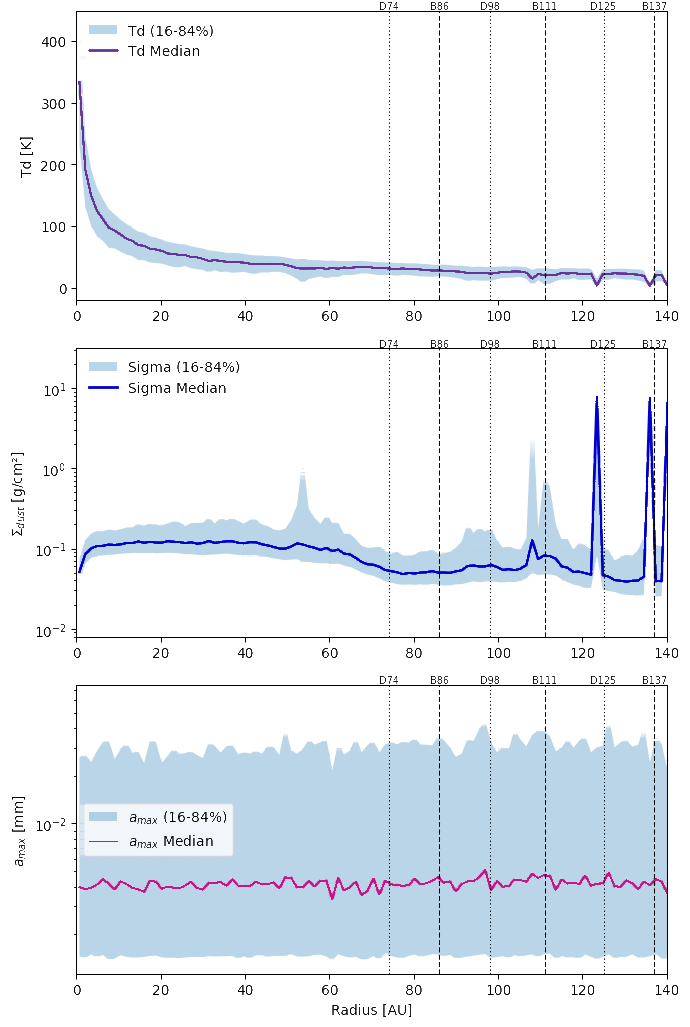}
    \caption{Physical properties ($T_d,\ \Sigma_{\text{dust}},\ \alpha_{\text{max}}$) of DoAr 25 in $1\mu m-1mm$ grain sizes calculated with a physical temperature prior. The bright rings and gaps are again noted 
following the convention by \citep{huang2018disk}.}
        \label{fig:doar_small_prior}
\end{figure}

\begin{figure*}[htbp!]
    \centering
    \subfloat[DoAr 25]{
        \includegraphics[width=0.47\linewidth]{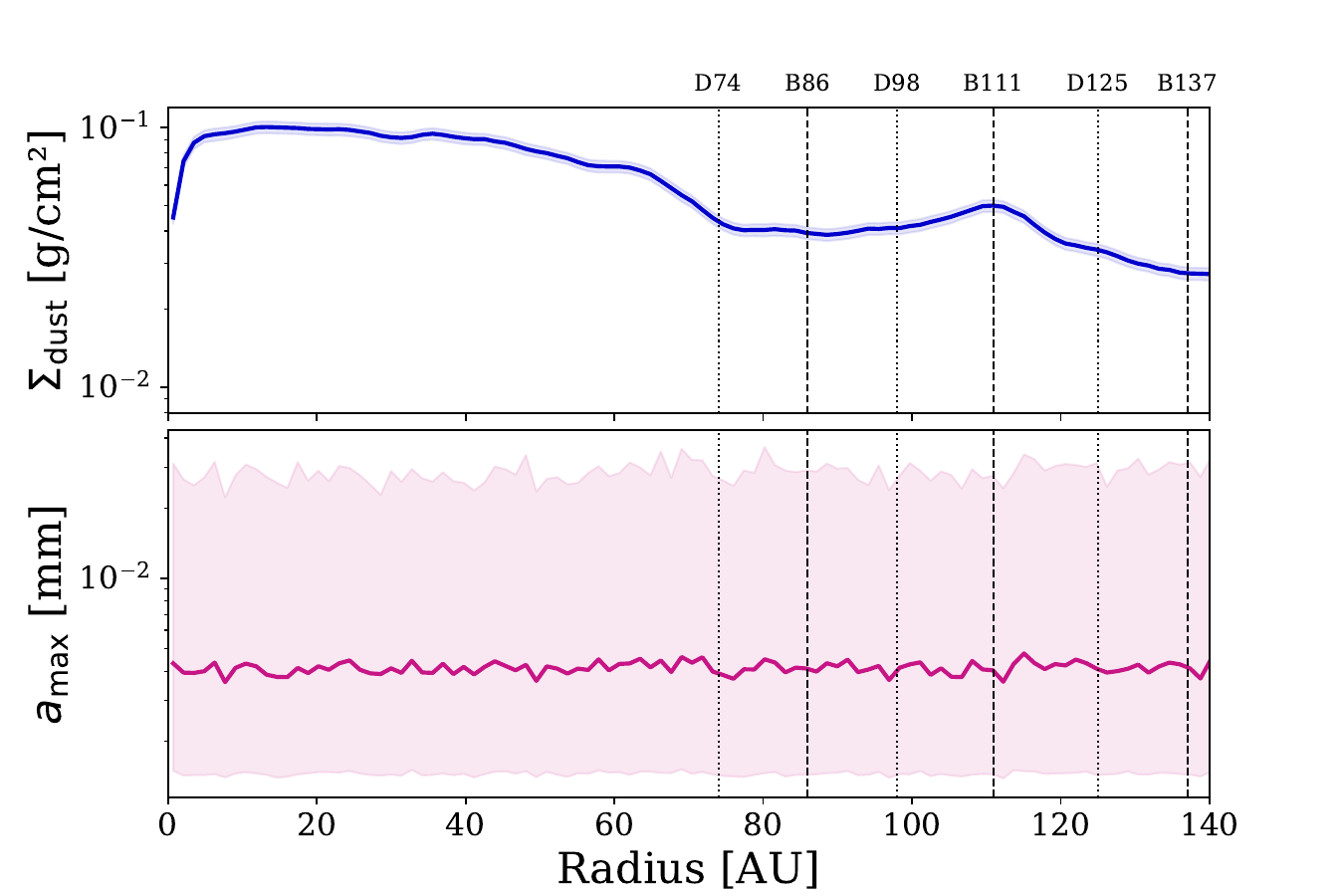}%
        \label{fig:properties_doar25_small}%
    }%
    \hfill
    \subfloat[CY Tau]{
        \includegraphics[width=0.47\linewidth]{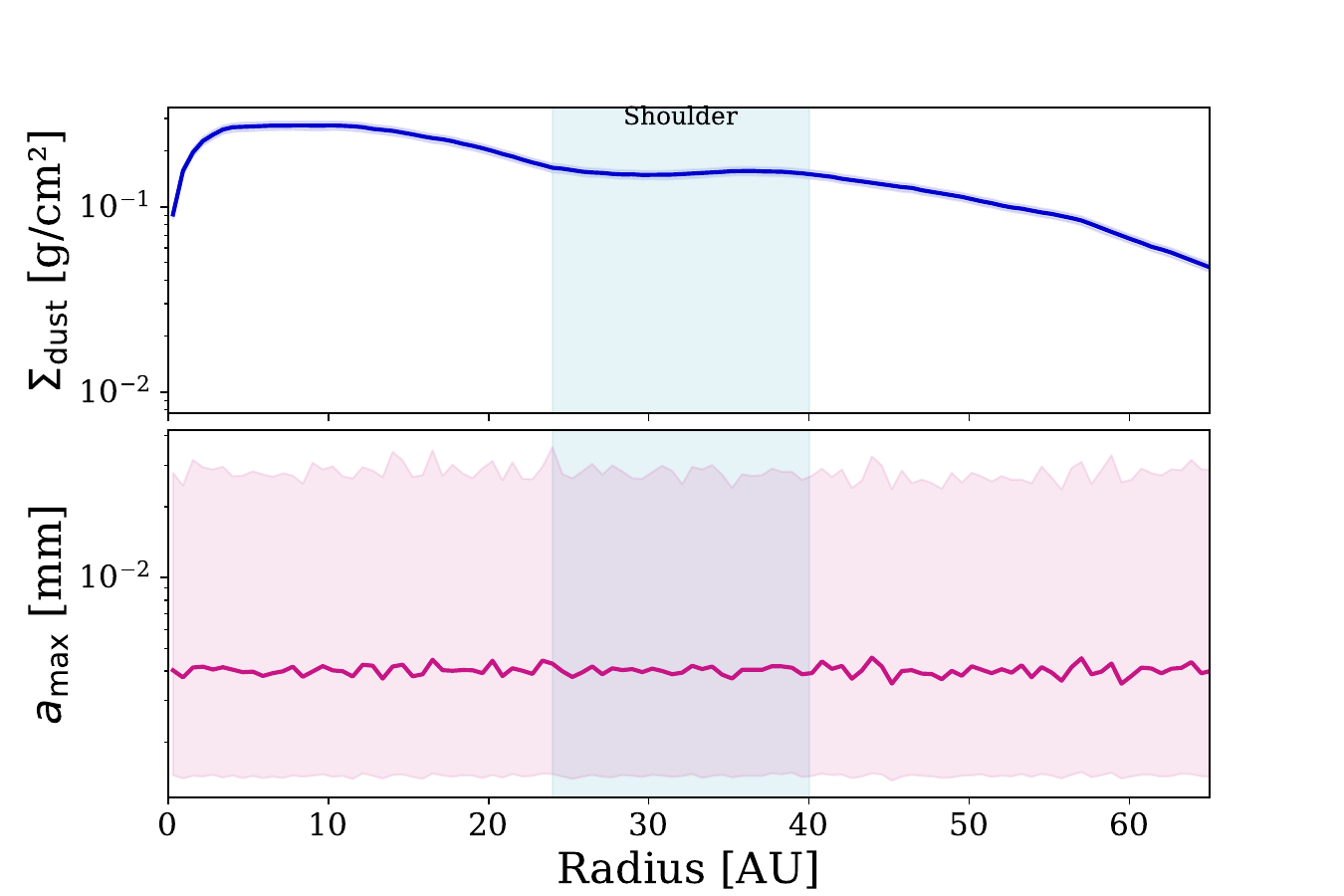}%
        \label{fig:properties_CYTau_small}%
    }
    \caption{Physical properties of DoAr 25 and CY Tau derived from radiative transfer modeling for $1\ \mu\mathrm{m}-1\ \mathrm{mm}$ grain sizes with a power-law temperature prior. Surface density is shown in blue and maximum emitting grain size in pink. Panel (a) shows DoAr 25 with bright rings and gaps noted following \cite{huang2018disk}. Panel (b) shows CY Tau with the shoulder feature indicated in light blue.}
    \label{fig:properties_sidebyside}
\end{figure*}

\begin{figure*}[htbp!]
    \centering
    \includegraphics[width=\textwidth, height=0.55\linewidth]{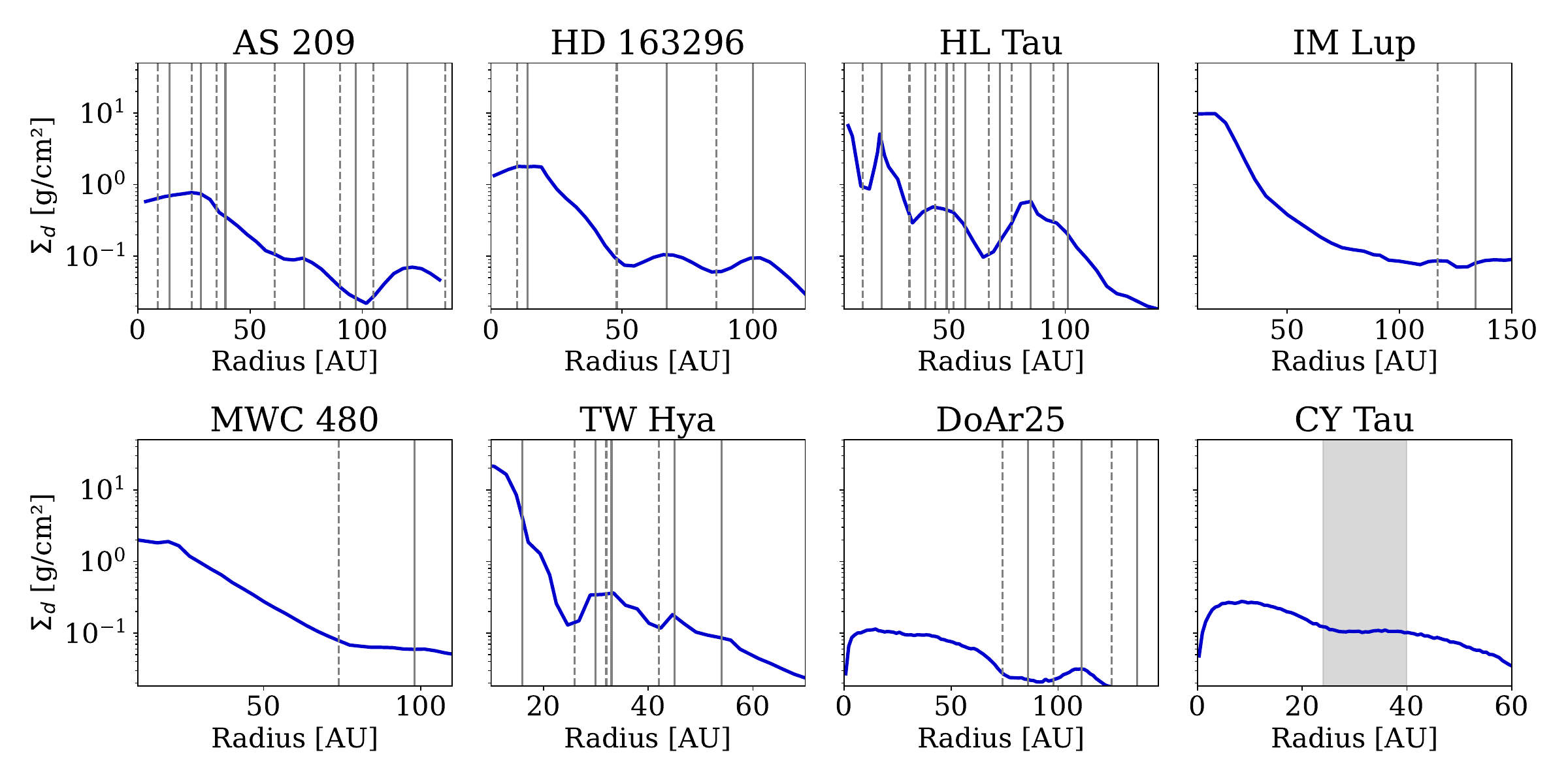}
    \caption{Comparison of dust surface density $\Sigma_d$ radial profiles of our disks (CY Tau and DoAr 25) with a sample of full disks. The locations of the rings are noted on the plots with the solid lines and the locations of the gaps with the dashed lines. CY Tau's shoulder feature is noted with a shaded area.}
    \label{fig:regular-sigma}
\end{figure*}

\begin{figure*}[htbp!]
    \centering
    \includegraphics[width=\textwidth, height=0.55\linewidth]{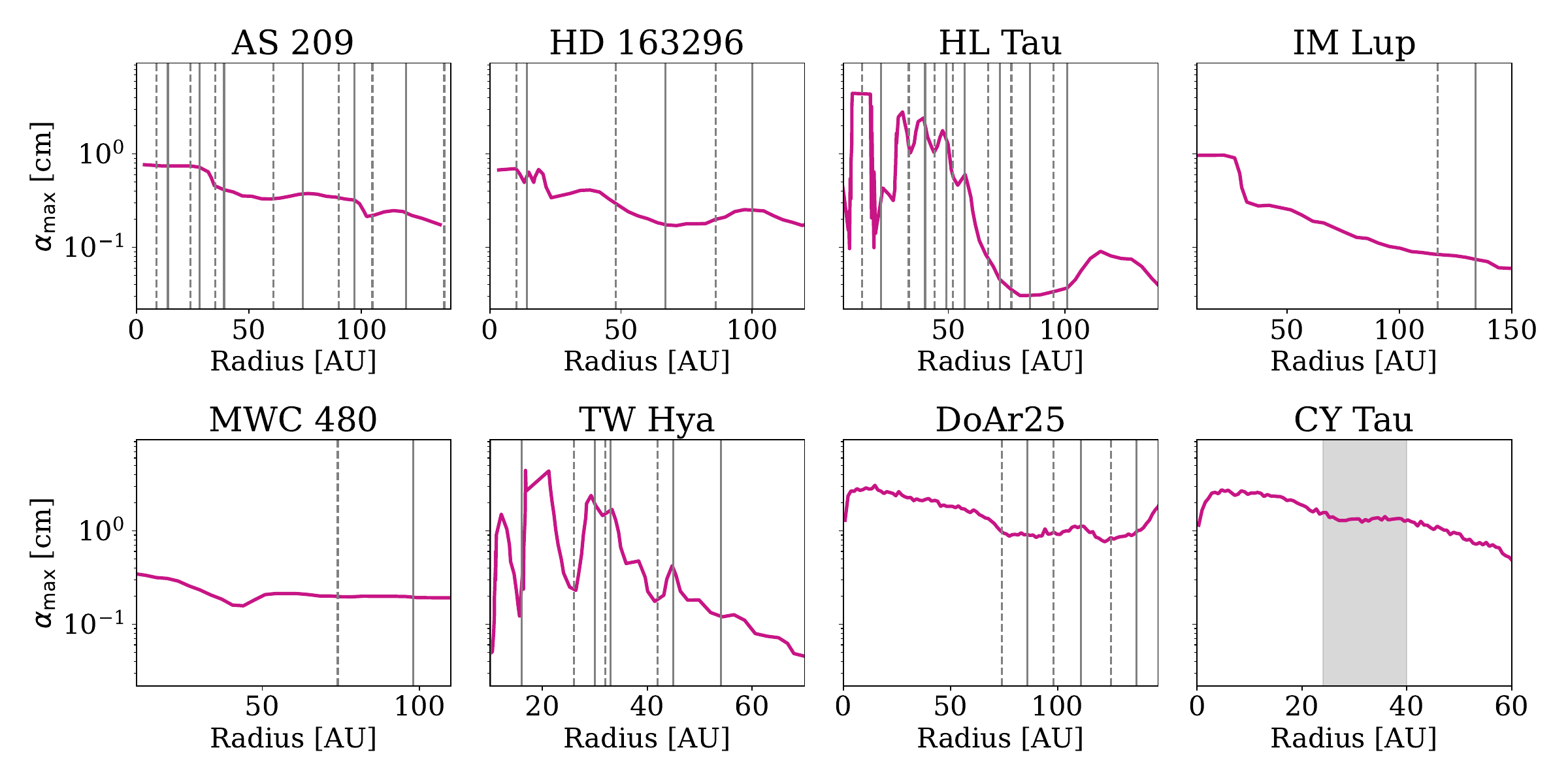}
    \caption{Comparison of maximum emitting grain size $\alpha_{\text{max}}$ radial profiles of our disks (CY Tau and DoAr 25) with a sample of full disks. The locations of the rings are noted on the plots with the solid lines and the locations of the gaps with the dashed lines. CY Tau's shoulder feature is noted with a shaded area.}
    \label{fig:regular-alpha}
\end{figure*}

\begin{figure*}[htbp!]
    \centering
    \includegraphics[width=\textwidth, height=0.55\linewidth]{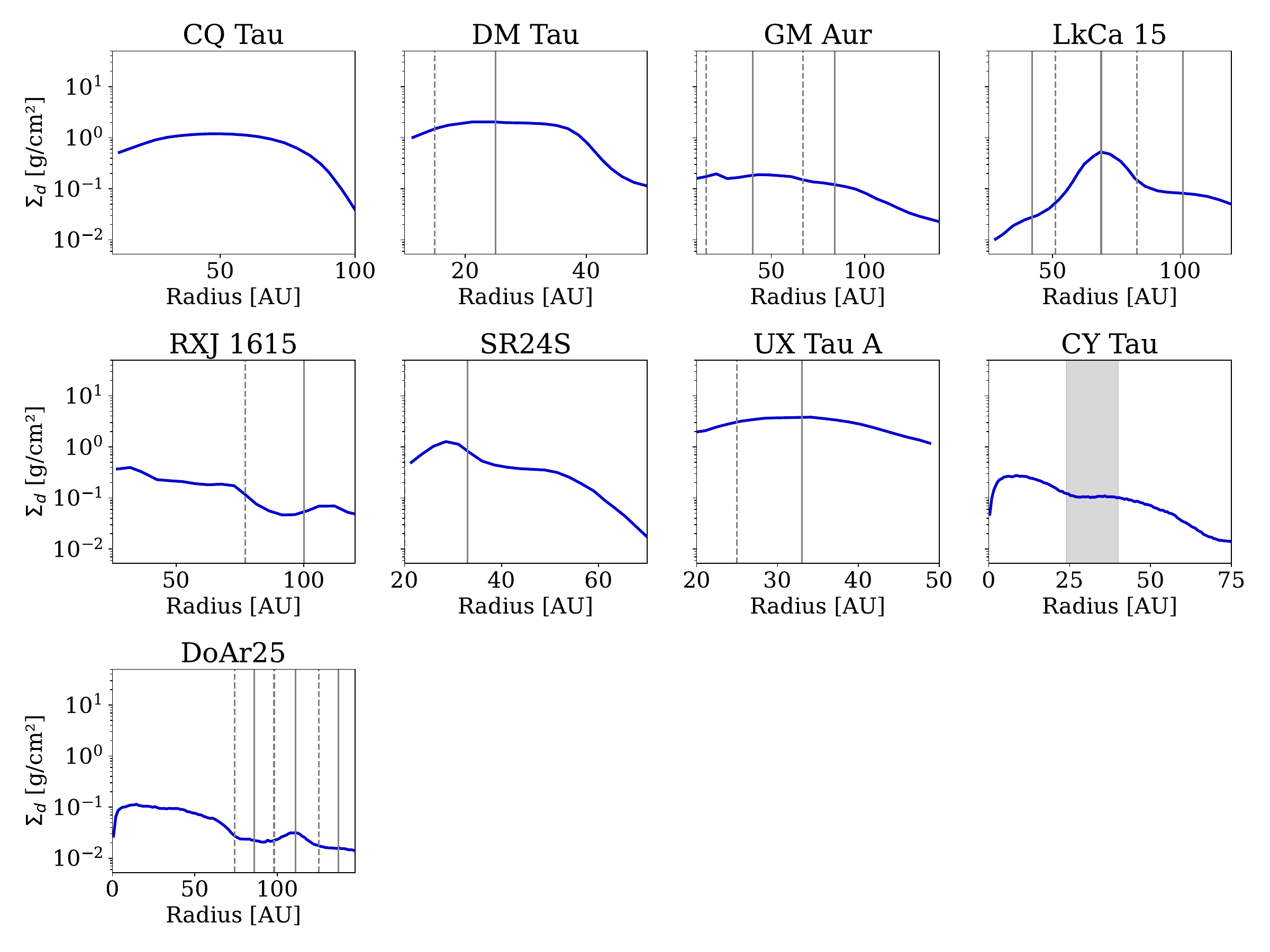}
    \caption{Comparison of dust surface density $\Sigma_d$ radial profiles of our disks (CY Tau and DoAr 25) with a sample of transition disks. The locations of the rings are noted on the plots with the solid lines and the locations of the gaps with the dashed lines. CY Tau's shoulder feature is noted with a shaded area.}
    \label{fig:transition-sigma}
\end{figure*}

\begin{figure*}[htbp!]
    \centering
    \includegraphics[width=\textwidth, height=0.55\linewidth]{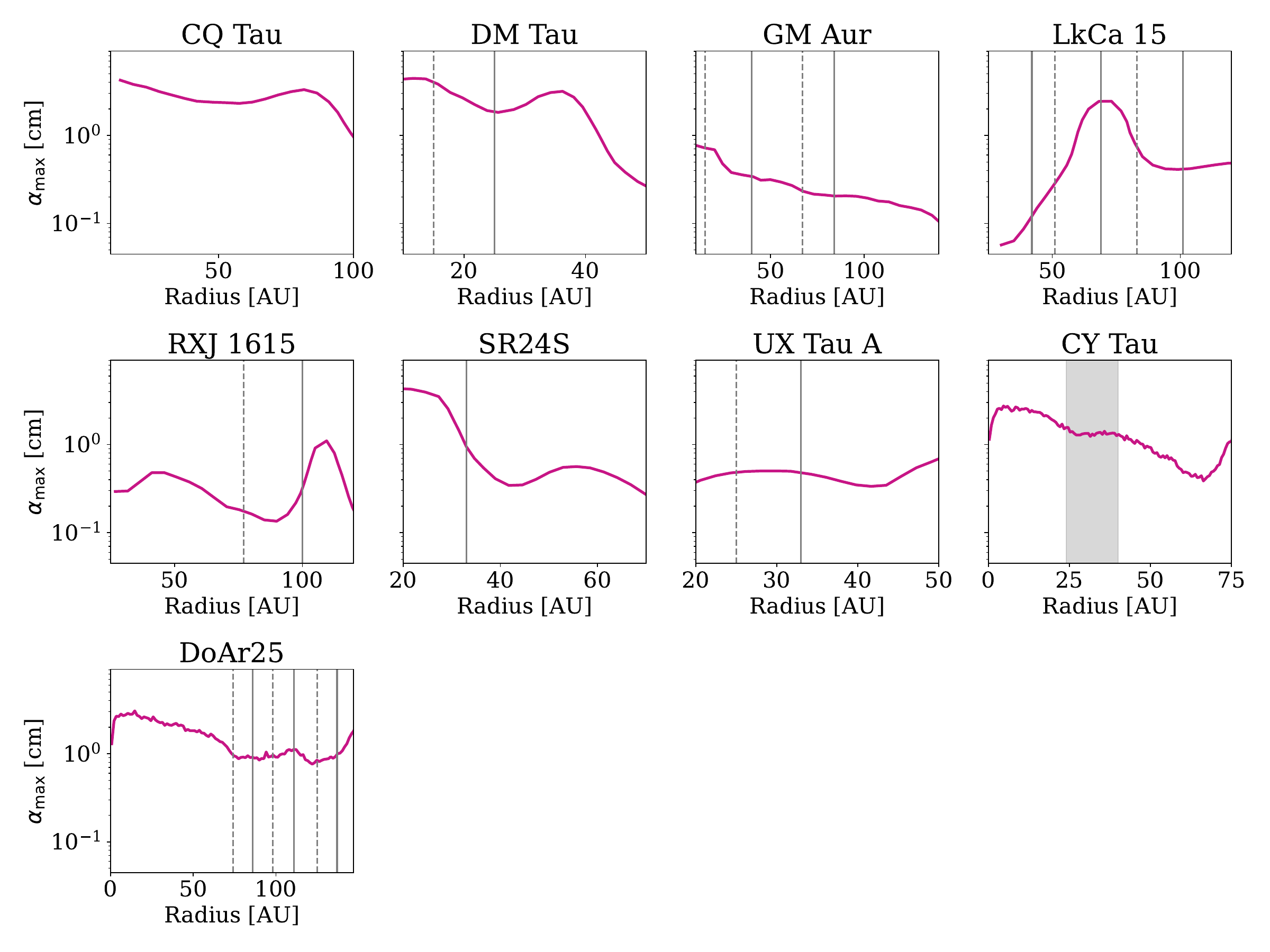}
    \caption{Comparison of maximum emitting grain size $\alpha_{\text{max}}$ radial profiles of our disks (CY Tau and DoAr 25) with a sample of transition disks. The locations of the rings are noted on the plots with the solid lines and the locations of the gaps with the dashed lines. CY Tau's shoulder feature is noted with a shaded area. The grain properties are derived from the cavity radius outward, as given in Table~\ref{tab:disk_summary}.}
    \label{fig:transition-alpha}
\end{figure*}

\begin{figure*}
    \centering
    \includegraphics[width=1\linewidth, height=0.55\linewidth]{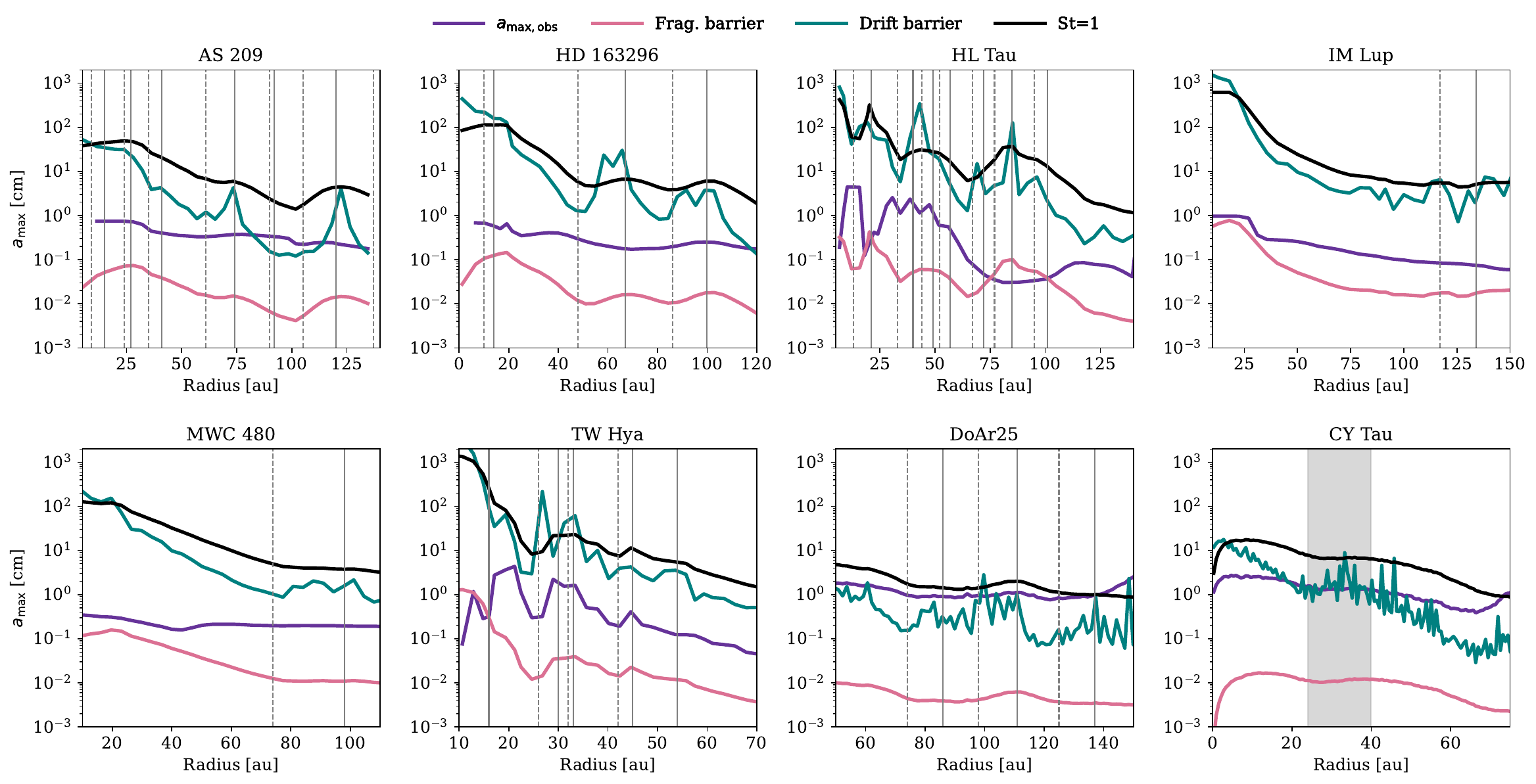}
    \caption{Radial profiles of the maximum grain size inferred from the radiative transfer modeling ($a_{\rm{max,obs}}$ with purple) compared with the theoretical fragmentation ($a_{\rm{frag}}$ with pink) and drift ($a_{\rm{drift}}$ with teal) barriers for the sample disks. The barriers are calculated assuming $v_{\rm{frag}}=1$ m/s$^{-1}$, $\alpha=10^{-3}$ and g2d=100. The black line indicates the Stokes number limit St=1.}
    \label{fig:frag_barriers_v1}
\end{figure*}

\begin{table}[htbp!]
\centering
\caption{Dust and central star masses of the sample disks in Earth and solar masses respectively.}

\begin{tabular}{lcc}
\hline\hline
Disk & $M_{\mathrm{dust}}$ [$M_\oplus$] & M$_*$ [M$_\odot$]\\
\hline
\multicolumn{3}{c}{\textbf{Literature-Transition disks}}\\
\specialrule{1.5pt}{0pt}{0pt}
CQ Tau     & 804 & 1.63 \\
DM Tau     & 305 & 0.39 \\
GM Aur     & 243 & 1.1 \\
LkCa 15    & 247 & 1.32 \\
RXJ 1615   & 232 & 1.1 \\
SR24S      & 185 & 0.87 \\
UX Tau A   & 663 & 1.4 \\
\hline
\multicolumn{3}{c}{\textbf{Literature-Full disks\tablefootmark{a}}}\\
\specialrule{1.5pt}{0pt}{0pt}
AS 209     & 249 & 1.2 \\
HD 163296  & 276 & 2.0 \\
HL Tau\tablefootmark{b} & 667 & 1.7 \\
IM Lup     & 1212 & 1.1 \\
MWC 480    & 396 & 2.1 \\
TW Hya     & 290 & 0.6 \\
\hline
\multicolumn{3}{c}{\textbf{This Work}}\\
\specialrule{1.5pt}{0pt}{0pt}
DoAr 25    & 213 & 0.65 \\
CY Tau     & 63 & 0.3\\
\hline
\end{tabular}
\caption*{\textbf{Note:} Here we present the values for the stellar masses and dust masses used to create Fig.~\ref{fig:dust_mass_star_mass}. 
\\ \textbf{References:} See Table~\ref{tab:disk_summary} and
\\ \tablefootmark{a} \cite{2024ApJ...974..306S} \\ \tablefootmark{b} \cite{guerra2024into}}
\label{tab:dust_masses}
\end{table}

\end{appendix}
\end{document}